\documentclass[aps,reprint,twocolumn,superscriptaddress,citeautoscript,amsmath,amssymb,floatfix,longbibliography,prb,]{revtex4-2}
\usepackage{graphicx}
\usepackage{times}
\usepackage{bm}
\usepackage{color}
\usepackage{gensymb}
\usepackage{dcolumn}
\usepackage{xfrac}

\usepackage{lipsum}

\newcommand{\TMn}[0]{$T_{\text{N}_{\text{Mn}}}$}
\newcommand{\TEuo}[0]{$T_{\text{N}_{\text{Eu1}}}$}
\newcommand{\TEut}[0]{$T_{\text{N}_{\text{Eu2}}}$}

\newcolumntype{d}[1]{D{.}{.}{#1}}

\begin{document}
\title{Canted antiferromagnetic phases in the layered candidate Weyl material EuMnSb$_{2}$ }

\author{J.~M.~Wilde}
\affiliation{Ames Laboratory, U.S. DOE, Iowa State University, Ames, Iowa 50011, USA}
\affiliation{Department of Physics and Astronomy, Iowa State University, Ames, Iowa 50011, USA}

\author{S.~X.~M.~Riberolles}
\affiliation{Ames Laboratory, U.S. DOE, Iowa State University, Ames, Iowa 50011, USA}

\author{Atreyee Das}
\affiliation{Ames Laboratory, U.S. DOE, Iowa State University, Ames, Iowa 50011, USA}
\affiliation{Department of Physics and Astronomy, Iowa State University, Ames, Iowa 50011, USA}

\author{Y. Liu}
\altaffiliation[Present Address:  ]{Crystal Growth Facility, Institute of Physics, \'{E}cole Polytechnique F\'{e}d\'{e}rale de Lausanne, CH-1015 Lausanne, CH}
\affiliation{Ames Laboratory, U.S. DOE, Iowa State University, Ames, Iowa 50011, USA}

\author{T.~W.~Heitmann}
\affiliation{University of Missouri Research Reactor, University of Missouri, Columbia, Missouri 65211, USA}
\affiliation{Department of Physics and Astronomy, University of Missouri, Columbia, Missouri 65211, USA}

\author{X.~Wang}
\affiliation{Neutron Scattering Division, Oak Ridge National Laboratory, Oak Ridge, TN 37831, USA}

\author{W.~E.~Straszheim}
\affiliation{ Materials Analysis and Research Laboratory, Office of Biotechnology,
	Iowa State University, Ames, Iowa 50011, USA}

\author{S.~L.~Bud`ko}
\affiliation{Ames Laboratory, U.S. DOE, Iowa State University, Ames, Iowa 50011, USA}
\affiliation{Department of Physics and Astronomy, Iowa State University, Ames, Iowa 50011, USA}

\author{P.~C.~Canfield}
\affiliation{Ames Laboratory, U.S. DOE, Iowa State University, Ames, Iowa 50011, USA}
\affiliation{Department of Physics and Astronomy, Iowa State University, Ames, Iowa 50011, USA}

\author{A.~Kreyssig}
\altaffiliation[Present Address:  ]{Ruhr-Universit\"{a}t Bochum, Universit\"{a}tsstra{\ss}e 150 44801 Bochum, DE}
\affiliation{Ames Laboratory, U.S. DOE, Iowa State University, Ames, Iowa 50011, USA}
\affiliation{Department of Physics and Astronomy, Iowa State University, Ames, Iowa 50011, USA}

\author{R.~J.~McQueeney}
\affiliation{Ames Laboratory, U.S. DOE, Iowa State University, Ames, Iowa 50011, USA}
\affiliation{Department of Physics and Astronomy, Iowa State University, Ames, Iowa 50011, USA}

\author{D.~H.~Ryan}
\email{dominic@physics.mcgill.ca}
\affiliation{Physics Department and Centre for the Physics of Materials,
	McGill University, 3600 University Street, Montreal, Quebec, H3A 2T8, CA}

\author{B.~G.~Ueland}
\email{bgueland@ameslab.gov, bgueland@gmail.com}
\affiliation{Ames Laboratory, U.S. DOE, Iowa State University, Ames, Iowa 50011, USA}
\affiliation{Department of Physics and Astronomy, Iowa State University, Ames, Iowa 50011, USA}

\date{\today}

\begin{abstract}
EuMnSb$_2$ is a candidate topological material which can be tuned towards a Weyl semimetal, but  there are differing reports for its antiferromagnetic (AFM) phases. The coupling of bands dominated by pure Sb layers hosting topological fermions to Mn and Eu magnetic states provides a potential path to tune the topological properties.  Here we present  single-crystal neutron diffraction, magnetization, and heat capacity data as well as polycrystalline $^{151}$Eu M\"ossbauer data which show that three AFM phases exist as a function of temperature, and we present a detailed analysis of the magnetic structure in each phase.  The Mn magnetic sublattice orders into a C-type AFM structure below \TMn$=323(1)$~K with the ordered Mn magnetic moment $\bm{\mu_{\text{Mn}}}$ lying perpendicular to the layers. AFM ordering of the Eu sublattice occurs below \TEuo$=23(1)$~K with the ordered Eu magnetic moment  $\bm{\mu}_{\text{Eu}}$ canted away from the layer normal and $\bm{\mu_{\text{Mn}}}$ retaining its higher temperature order.  $\bm{\mu}_{\text{Eu}}$ is ferromagnetically aligned within each Eu layer but exhibits a complicated AFM layer stacking.  Both of these higher temperature phases are described by magnetic space group (MSG) $Pn^{\prime}m^{\prime}a^{\prime}$ with the chemical and magnetic unit cells having the same dimensions.    Cooling below \TEut$=9(1)$~K reveals a third AFM phase where $\bm{\mu_{\text{Mn}}}$ remains unchanged but $\bm{\mu}_{\text{Eu}}$ develops an additional substantial in-plane canting.  This phase has MSG $P11\frac{2_1}{a^{\prime}}$.   We also find some evidence of short-range magnetic correlations associated with the Eu between $12~\text{K} \alt T \alt 30~\text{K}$.  Using the determined magnetic structures, we postulate the signs of nearest-neighbor intralayer and interlayer exchange constants and the magnetic anisotropy within a general Heisenberg-model.  We then discuss implications of the various AFM states in EuMnSb$_2$ and their potential for tuning topological properties.

\end{abstract}

\maketitle

\section{Introduction}
Topological semimetals offer exotic physical properties such as chiral-charge pumping and linear-negative longitudinal magnetoresistance associated with relativistic Weyl fermions \cite{Hasan_2010,Bansil_2016,Armitage_2018,Vanderbilt_2018,Lv_2021}.  These properties arise from nontrivial topology of the bulk and surface electronic-band structures, which are intimately connected to the underlying symmetry of the crystal lattice and any magnetic order.  Thus, modifying magnetic order can offer control of topological properties.

Relativistic (nearly massless) Dirac fermions generally occur in the linearly dispersing electronic bands forming Dirac cones \cite{Bansil_2016,Vanderbilt_2018, Armitage_2018}. Surface-Dirac cones with nodes at or near the Fermi energy $E_{\text{F}}$ are of particular interest because they can give rise to topologically induced transport properties \cite{Hasan_2010, Lv_2021}. Further, topological materials with a net magnetization $M$ that can be readily tuned by temperature or external fields are especially sought after because they can offer direct control of Dirac surface states \cite{Vanderbilt_2018}.  For example, the presence of a net $M$ in a topological-crystalline insulator can result in the gapping of a surface-Dirac cone:  gapless surface-Dirac cones are associated with chiral-locked dissipationless surface conductivity or edge states whereas gapped surface-Dirac cones lead to quantum-anomalous-Hall type conductivity \cite{Riberolles_2021}.

The emergence of certain magnetic order or application of a strong enough magnetic field $H$ can lift the spin degeneracy of the bulk-Dirac cones in a semimetal \cite{Soh_2021}.  This creates two Weyl nodes for each Dirac node and the associated relativistic Weyl fermions.  The two Weyl nodes have opposite chirality (i.e.\ $+$ or $-$ chirality) and can be viewed as chiral monopoles (sources and sinks of Berry curvature) connected by surface Fermi arcs.  An unequal number of $+$ or $-$ chiral charges created through, for example, the application of parallel electric and magnetic fields results in the chiral anomaly which gives rise to important topologically induced properties such as linear-negative-longitudinal magnetoresistance \cite{Lv_2021}.

Examples of Weyl semimetals (WSMs) have been observed for crystals with broken inversion symmetry $\mathcal{I}$ (e.g.\ TaAs \cite{Lv_2015} and NbAs \cite{Xu_2015}).  However, very few examples of materials with broken time-reversal symmetry $\mathcal{T}$ exist, and magnetic  ordering offers a way to break $\mathcal{T}$.  Searches for such broken $\mathcal{T}$ WSMs have included $A$Mn$X_2$, $A =$~Ca, Sr, Ba, Yb, or Eu and $X=$~Sb or Bi, with magnetic moments due to the Mn and Eu and highly anisotropic Dirac cones within the Sb and Bi square or zig-zag nets \cite{Feng_2014, Liu_2017,Liu_2019}. As compared to tetragonal $112$ compounds, the orthorhombic variants with $A =$~Eu, Yb, and Sr are particularly interesting because ferromagnetic (FM) canting of the ordered magnetic moment $\bm{\mu}$ can occur without further reduction of the underlying chemical symmetry.  Such canting can lift the degeneracy of the Dirac cones to form Weyl nodes. 

\begin{figure*}[]
	\centering
	\includegraphics[width=1.0\linewidth]{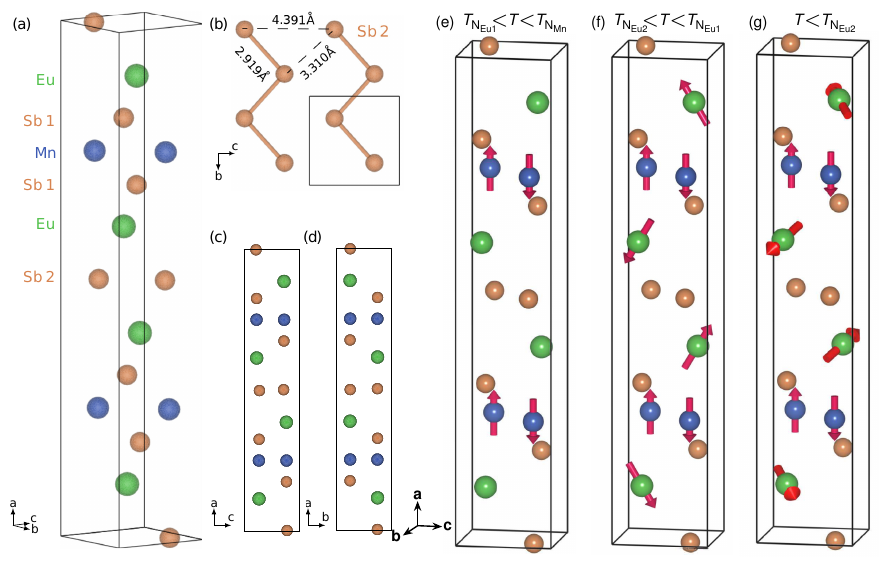}
	\caption{\label{Fig:crys_struct} (a) The chemical structure of EuMnSb$_2$ with $a=22.4958$~\AA, $b=4.3758$~\AA, and $c=4.3908$~\AA.  The orthorhombic unit cell is drawn with gray lines, and the two Sb sites are labeled Sb~$1$ and Sb~$2$. (b) The Sb~$2$ layer as viewed by looking down the $\mathbf{a}$ axis.  (c),(d) Views of the chemical unit cell looking down $\mathbf{b}$ (c) and up $\mathbf{c}$ (d). (e), (f), and (g) show the antiferromagnetic order occurring at temperatures of \TEuo$<T<$~\TMn, \TEut$<T<$~\TEuo, and $T<$~\TEut, respectively, where  \TMn$=323(1)$~K, \TEuo$=23(1)$~K, and \TEut$=9(1)$~K.  These diagrams were made using \textsc{vesta} \cite{Momma_2011}.}
\end{figure*}

EuMnSb$_2$ is a particularly compelling candidate Weyl semimetal because  it contains two magnetic species and the slightly orthorhombic unit cell permits the existence of a net $M$ canted away from the layer normal without the magnetic domain formation required for the tetragonal $112$s.  Figure~\ref{Fig:crys_struct} shows the orthorhombic chemical-unit cell for EuMnSb$_2$ which has space group $Pnma$ and lattice parameters of $a=22.4958(1)$~\AA, $b=4.3758(1)$~\AA, and $c=4.3908(1)$~\AA\ at a temperature of $T=50$~K.  Such $A$Mn$X_2$ compounds generally contain repeating $A$-$X1$-Mn-$X1$-$A$ blocks along $\mathbf{a}$ that are separated by $X2$ layers hosting topological fermions.  Here, $X1$ and $X2$ denote $X$ ions on different crystallographic sites.  For the orthorhombic unit cell, the $X2$ layers are not square, as found in the tetragonal $112$s, but are made up of the zig-zag chains shown in Fig.~\ref{Fig:crys_struct}(b).

 The occurrence of both topological fermions and magnetic ions provides a potential path to tune the topological properties via magnetic order.  Indeed,  Hall resistivity data for antiferromagnetic (AFM) EuMnBi$_2$ show the occurrence of the half-integer-quantum-Hall effect where control of the Eu magnetic sublattice via a magnetic field is reported to suppress the interlayer coupling between Bi layers hosting Dirac fermions \cite{Masuda_2016}.  For EuMnSb$_2$, angle-resolved-photoemission-spectroscopy (ARPES) data show Dirac like linear dispersion near $E_{\text{F}}$, and magnetotransport data indicate that exchange induced effects of  the Eu magnetism affect the electronic transport \cite{Soh_2019}. 

 The magnetic ordering of  EuMnSb$_2$ has been previously studied by neutron diffraction using both powder \cite{Soh_2019} and single-crystal samples \cite{Gong_2020, Zhang_2022}, however, Ref.~\onlinecite{Zhang_2022} reports a tetragonal rather than orthorhombic chemical-unit cell.  All three studies find two AFM transitions, one at a N\'eel temperature of $T_{\text{N}_{\text{Mn}}}=350(2)$~K \cite{Soh_2019} ($T_{\text{N}_{\text{Mn}}}\approx346$~K \cite{Gong_2020}, $T_{\text{N}_{\text{Mn}}}\approx340$~K \cite{Zhang_2022}) due to ordering of the Mn and one at a N\'eel temperature of $T_{\text{N}_{\text{Eu}}}=21(1)$~K \cite{Soh_2019} ($T_{\text{N}_{\text{Eu}}}\approx21$~K \cite{Gong_2020}, $T_{\text{N}_{\text{Eu}}}\approx22$~K \cite{Zhang_2022}) owing to ordering of the Eu.  For the Mn AFM order the studies find the C-type magnetic structure shown in Fig.~\ref{Fig:crys_struct}(e), with neighboring Mn spins within a layer being antiferromagnetically oriented and FM alignment of neighboring Mn spins along the direction perpendicular to the Sb layers.  The studies concluded that the ordered Mn magnetic moment $\bm{\mu}_{\text{Mn}}$ lies perpendicular to the Sb layers.and that the Mn sublattice retains this magnetic structure upon cooling through $T_{\text{N}_{\text{Eu}}}$ \cite{Soh_2019, Gong_2020}.  Resistivity versus temperature data show a maximum at $T_{\text{N}_{\text{Eu}}}$ and either metallic \cite{Soh_2019} or semiconducting \cite{Yi_2017} behavior below $T_{\text{N}_{\text{Eu}}}$.

The results from the powder \cite{Soh_2019} and single-crystal \cite{Gong_2020, Zhang_2022} neutron diffraction studies differently describe the magnetic structure of the Eu sublattice below $T_{\text{N}_{\text{Eu}}}$.  The powder study finds A-type AFM order with FM Eu layers stacked antiferromagnetically along $\mathbf{a}$ and an ordered Eu magnetic moment $\bm{\mu}_{\text{Eu}}$ lying along $\mathbf{c}$ \cite{Soh_2019}.  Ref.~\onlinecite{Gong_2020} finds a canted AFM structure with $\bm{\mu}_{\text{Eu}}$ lying in the $\mathbf{ac}$ plane .  The structure  consists of FM Eu layers with a staggered $\mathbf{a}$ component of $\bm{\mu}_{\text{Eu}}$ in neighboring layers. The relative orientation of the $\mathbf{c}$ component of $\bm{\mu}_{\text{Eu}}$ has a $\rightarrow\rightarrow\leftarrow\leftarrow$ pattern along $\mathbf{a}$, as shown in Fig.~\ref{Fig:crys_struct}(f). Ref.~\onlinecite{Zhang_2022} finds an AFM structure with a  $\leftarrow\rightarrow\rightarrow\leftarrow$ pattern along $\mathbf{c}$, but $\bm{\mu}_{\text{Eu}}$ lying solely along $\mathbf{a}$.  Note that in the tetragonal setting of Ref.~\onlinecite{Zhang_2022} $\mathbf{c}$ is perpendicular to the Eu, Sb$1$-Mn-Sb$1$, and Sb$2$ layers.  There is zero $M$ associated with the determined orders, and  Ref~\onlinecite{Gong_2020} found that $\mu_{\text{Mn}}=4.5(6)~\mu_{\text{B}}$ and $\mu_{\text{Eu}}=5.9(8)~\mu_{\text{B}}$ at $T=7$~K.  Ref.~\onlinecite{Zhang_2022} reports $\mu_{\text{Mn}}=4.6(2)~\mu_{\text{B}}$ and $\mu_{\text{Eu}}=5.2(4)~\mu_{\text{B}}$ at $5$~K.  Notably, the value of $\mu_{\text{Eu}}$ is less than the expected value of $7~\mu_{\text{B}}$ for Eu$^{2+}$.

Here, we report magnetization, heat capacity, $^{151}$Eu M\"ossbauer spectroscopy, and single-crystal neutron diffraction results for EuMnSb$_2$.  Our data are consistent with a canted AFM phase below $T_{\text{N}_{\text{Eu}}}$ and we present results showing a third AFM phase occurring below \TEut$=9(1)$~K.  Our detailed single-crystal neutron diffraction study has determined the magnetic structure in each of the three AFM phases, which are shown in Figs.~\ref{Fig:crys_struct}(e)--\ref{Fig:crys_struct}(g).  Based on these results, we postulate the sign of nearest-neighbor (NN) exchange interactions and magnetic anisotropy for a generic Heisenberg model and make comparisons with EuMnBi$_2$ and SrMnSb$_2$, as well as other reports for EuMnSb$_2$.  We also discuss the possible influences of disorder, the geometric arrangement of ions in the $X2$ layers, and the magnetic order on the topological properties.

\section{Experiment}
Plate-like single crystals of EuMnSb$_{2}$ were flux grown and screened by x-ray diffraction. The growth of EuMnSb$_2$ single crystals is similar to the growth of SrMnSb$_2$ single crystals described in detail in Ref.~[\onlinecite{Liu_2019}]. Europium pieces, manganese powder, and antimony chunks were weighed at a molar ratio of Eu:Mn:Sb~$=1$:$1$:$10$ and loaded into an alumina crucible in a glovebox with an argon atmosphere. The alumina crucible was sealed in an evacuated quartz tube with a backfilling of $300$~mbar of argon gas and then heated up to $T=1273$~K. After a dwell time of $12$ hours, the tube was slowly cooled to $873$~K at a rate of $3~\text{K}/\text{h}$. Plate-like single crystals with masses of $m\alt40$~mg were cleaved from the matrix.  X-ray diffraction data for ground single crystals were consistent with previous results for the chemical-unit cell although significant texture induced effects in the diffraction pattern prevented accurate determination of the stoichiometry. The composition of single crystals of EuMnSb$_2$ were quantitatively determined by Energy-Dispersive Spectroscopy (EDS)  as described in Sec.~\ref{Sec:EDS}.

Magnetization measurements were made down to $T=1.8$~K and up to $\mu_{0}H=7$~T using a Quantum Design, Inc., Magnetic Property Measurement System with a superconducting quantum interference device.  Measurements of the heat capacity at constant pressure $C_{\text{p}}$ were made down to $1.8$~K in a Quantum Design, Inc., Physical Property Measurement System using a standard semi-adiabatic heat-pulse technique.

$^{151}$Eu M\"ossbauer spectroscopy measurements were carried out using a $4$~GBq $^{151}$SmF$_3$ source driven in sine mode and calibrated using a standard $^{57}$Co\underline{Rh}/$\alpha$-Fe foil.  Isomer shifts are quoted relative to EuF$_3$ at ambient temperature. The sample consisted of ground single crystals and was cooled in a vibration-isolated-closed-cycle-He refrigerator with the sample in He exchange gas.  The hyperfine spectra were fitted at each temperature to a sum of Lorentzian lineshapes. The intensities and positions of the lines were derived from a full solution to the nuclear Hamiltonian \cite{Voyer_2006}. Spectra taken above $T=10$~K were also fitted using a model that derives a distribution of hyperfine fields by assuming an incommensurate magnetic structure with a sinusoidally modulated value for $\mu$ \cite{Bonville_2001, Maurya_2014}.

Single-crystal neutron diffraction measurements were made on the TRIAX triple-axis-neutron spectrometer at the University of Missouri Research Reactor and on the TOPAZ diffractometer at the Spallation Neutron Source, Oak Ridge National Laboratory. Measurements on TRIAX were made using a neutron wavelength of $\lambda=1.638$~\AA\ selected by a pyrolitic-graphite (PG) monochromator.  A PG analyzer was employed to reduce background scattering.  S\"oller-slit collimators with divergences of $60^{\prime}$-$60^{\prime}$-$80^{\prime}$-$80^{\prime}$ were inserted before the monochromator, between the  monochromator and sample, between the sample and analyzer, and between the analyzer and detector, respectively.  Higher resolution measurements were made using $60^{\prime}$-$60^{\prime}$-$20^{\prime}$-$20^{\prime}$  S\"oller-slit collimators. PG filters were inserted before and after the sample to reduce contamination by higher order neutron wavelengths.  

On TRIAX, the sample was mounted to an Al holder and placed inside of an Al can.  The can was filled with He exchange gas, sealed, and attached to the cold head of a closed-cycle-He refrigerator which allowed for measurements between $T=6$ and $340$~K. As explained below, crystal twinning permitted the simultaneous collection of data for both the $(h\,k\,0)$ and $(h\,0\,l)$ scattering planes, where $h$, $k$, and $l$ are Miller indices.  Corrections for neutron absorption were performed with \textsc{mag2pol} \cite{Qureshi_2019}.

TOPAZ is a time-of-flight neutron diffractometer which utilizes the wavelength-resolved-Laue-diffraction technique \cite{Coates_2018}. The same single-crystal sample used for the TRIAX experiments was connected to an Al holder and attached to the instrument's cryogenic goniometer which allowed for cooling down to $T=5$~K.  Measurements were made at $5$, $12$, and $50$~K for various orientations of the crystal with the $(h\,0\,l)$ plane horizontal.  Twinning was also observed during this experiment and allowed for the observation of peaks in the $(h\,k\,0)$ plane.  Corrections for neutron absorption were performed using \textsc{anvred} \cite{Schultz_1984} and data were analyzed using single-crystal refinements done with \textsc{jana} \cite{Duvsek_2001}.

\section{Results}

\subsection{Energy Dispersive Spectroscopy}\label{Sec:EDS}

EDS measurements were made using an Oxford Instruments Aztec EDS x-ray analyzer with an X-Max-80 detector mounted on an FEI Quanta-FEG scanning electron microscope. Samples were measured using a voltage of $15$~kV and a current of $\approx0.8$ nA. Spectra were collected for $30$~s at $19000$~counts~per~s. The Eu $L$ line series was used to quantify the amount of Eu with reference to a vitreous Eu standard. The Mn $K$ line series and the Sb $L$ line series were used with reference to pure element standards. All standards were internal to the Oxford software. Several other reference materials, e.g. Eu phosphate, were used to establish the accuracy of the results.

The average from $11$ spot measurements spanning $2$ different areas of a single crystal yielded ratios of $\text{Mn}/\text{Sb}=0.49(1)$, $\text{Eu}/\text{Sb}=0.51(1)$,  and $\text{Eu}/\text{Mn}=1.04(2)$ for the stoichiometry.  The measurement precision for the Eu content  is $1$ part out of $79$ ($1.3\%$). 

\subsection{Magnetization and Heat Capacity}

\begin{figure}[]
	\centering
	\includegraphics[width=1\linewidth]{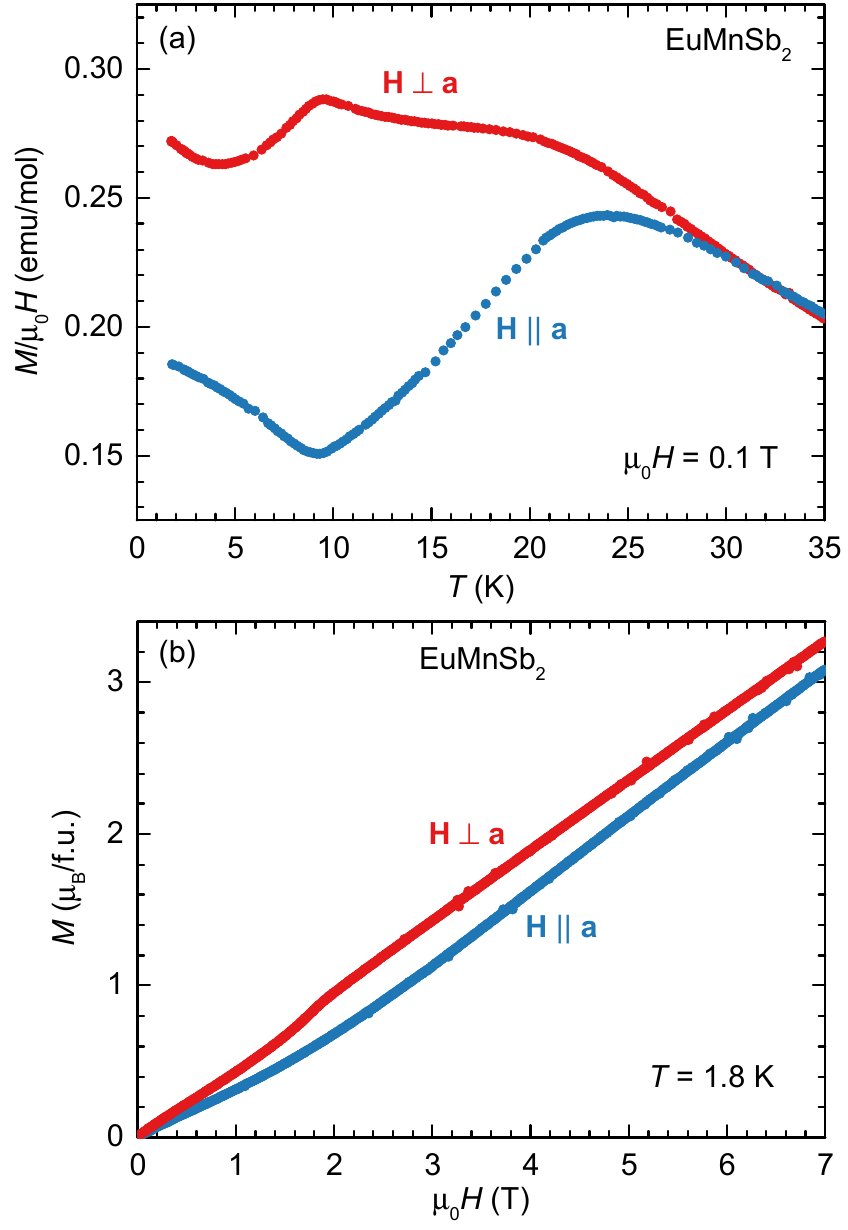}
	\caption{\label{Fig:Mag} (a) The magnetization divided by magnetic field versus temperature for $\mathbf{H}$ applied parallel or perpendicular to $\mathbf{a}$. (b) The magnetization versus magnetic field at $T=1.8$~K for $\mathbf{H}$ applied parallel or perpendicular to $\mathbf{a}$. f.u.\ stands for formula unit.}
\end{figure}

Figure~\ref{Fig:Mag}(a) shows $M/\mu_0H$ versus temperature for $\mu_0H=0.1$~T for $\mathbf{H}\parallel\mathbf{a}$ and $\mathbf{H}\perp\mathbf{a}$.  Changes in the slopes of the curves are evident for both field directions at $T\approx23$ and $9$~K which correspond to the AFM transitions at \TEuo\ and \TEut, respectively.  We show below via our neutron diffraction results that the magnetic ordering of the Mn sublattice does not change upon crossing \TEuo\ and \TEut, with \bm{$\mu_{\text{Mn}}}$ lying along $\mathbf{a}$ for $T<$~\TMn.  Rather, the two transition temperatures are associated with a paramagnetic (PM) to AFM transition for the Eu sublattice (\TEuo) and a change in the AFM structure of the Eu sublattice (\TEut).  With these facts in mind, examining the region between \TEuo\ and \TEut,  we see that $M/\mu_0H$ for $\mathbf{H}\parallel\mathbf{a}$ decreases with decreasing $T$ whereas $M/\mu_0H$ for $\mathbf{H}\perp\mathbf{a}$ gently increases.  For a typical AFM, these data would suggest that the ordered magnetic moment lies primarily along $\mathbf{a}$.  The data for $T<$~\TEut, on the other hand, suggest that the ordered moment has a significant component perpendicular to $\mathbf{a}$.  This is consistent with the AFM structures we report below where we find from neutron diffraction that $\bm{\mu_{\text{Eu}}}$ has its largest component lying along $\mathbf{a}$ for \TEuo~$>T>$~\TEut\ but its largest component lies perpendicular to $\mathbf{a}$ for $T<$~\TEut.

Figure~\ref{Fig:Mag}(b) shows $M(H)$ at $T=1.8$~K for $\mathbf{H}\parallel\mathbf{a}$ and $\mathbf{H}\perp\mathbf{a}$.  Both curves increase with increasing $H$ and show no signs of magnetic saturation up to $\mu_0H=7$~T.  The curves are similar to those previously reported \cite{Gong_2020}, with the $\mathbf{H}\perp\mathbf{a}$ curve lying above the $\mathbf{H}\parallel\mathbf{a}$ curve and the two curves adopting linear behavior above $\approx2$~T.  Both show a change in slope beginning at $\approx 1.5$~T.  For $\mathbf{H}\parallel\mathbf{a}$, the deviation may be associated with gradual rotation of $\bm{\mu}_{\text{Eu}}$ towards $\mathbf{a}$ because, as we show below, $\bm{\mu}_{\text{Mn}}$ lays only along $\mathbf{a}$ whereas $\bm{\mu}_{\text{Eu}}$ has components along all three crystallographic directions. One might expect a spin-flop of the Mn spins for $\mathbf{H}\parallel\mathbf{a}$ if $\bm{\mu}_{\text{Mn}}$ were reorienting and a previous report shows a step-like feature in $M(H)$ data at $2$~K for $\mathbf{H}\parallel\mathbf{a}$ which is attributed to a spin flop occurring at $1.5$~T \cite{Yi_2017}.  On the other hand, similar to  our data in Fig.~\ref{Fig:Mag}(b), data in Ref.~[\onlinecite{Gong_2020}] do not show a clear-cut indication for a spin flop.  For $\mathbf{H}\perp\mathbf{a}$, the change in slope near $1.5$~T may represent both $\bm{\mu}_{\text{Eu}}$ and $\bm{\mu}_{\text{Mn}}$ reorienting  towards $\mathbf{H}$.

\begin{figure}[]
	\centering
	\includegraphics[width=1\linewidth]{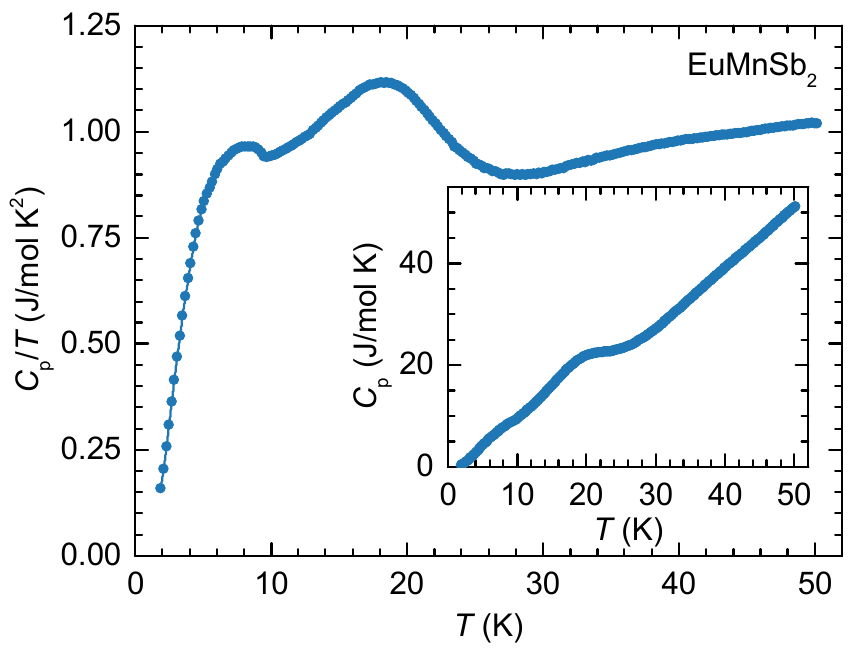}
	\caption{\label{Fig:Cp} The isobaric heat capacity divided by temperature versus temperature.  The inset shows the isobaric heat capacity versus temperature.}
\end{figure}

Figure~\ref{Fig:Cp} shows $C_{\text{p}}/T$ versus $T$ data and the inset shows $C_{\text{p}}(T)$.  Two peaks are observed: one broad peak corresponding to \TEuo\ and a weaker peak associated with \TEut.  Taken together with the features observed in $[M/\mu_0H](T)$, the peaks in $[C_{\text{p}}/T](T)$  indicate a change in entropy $S$ due to magnetic phase transitions at \TEuo\ and \TEut.  Our $[C_{\text{p}}/T](T)$ data spanning \TEuo\ are consistent with the broadness and the size of the peak shown in Ref.~[\onlinecite{Yi_2017}], which reports an estimated change in the magnetic component of $S$ of $\approx80\%$ of the value of $S_{\text{Eu}}=17.3$~J$/$mol-K expected for $S=7/2$ Eu$^{2+}$.  A less pronounced peak at \TEut\ is seen in Ref.~[\onlinecite{Yi_2017}], and we estimate that the small peak at \TEut\ in our $[C_{\text{p}}/T](T)$ data would contribute only an additional $S\approx0.1$~J$/$mol-K.  Nevertheless, the data we present below clearly indicate that the Eu sublattice undergoes successive magnetic transitions at temperatures corresponding to the peaks in $C_{\text{p}}(T)$.

AFM transitions typically are expected to yield much sharper peaks in $C_{\text{p}}(T)$ then those shown in Fig.~\ref{Fig:Cp}.  Broadening of these peaks could be due to disorder not detected by our EDS and x-ray measurements. However, as we discuss below, the $^{151}$Eu M\"ossbauer spectroscopy and neutron diffraction data show evidence for short-range magnetic correlations associated with the Eu sublattice which can also broaden the peaks in $C_{\text{p}}(T)$.

\subsection{$\bm{^{151}}$Eu M\"ossbauer spectroscopy}  \label{SubSec:Moss}

\begin{figure}[]
	\centering
	\includegraphics[width=1\linewidth]{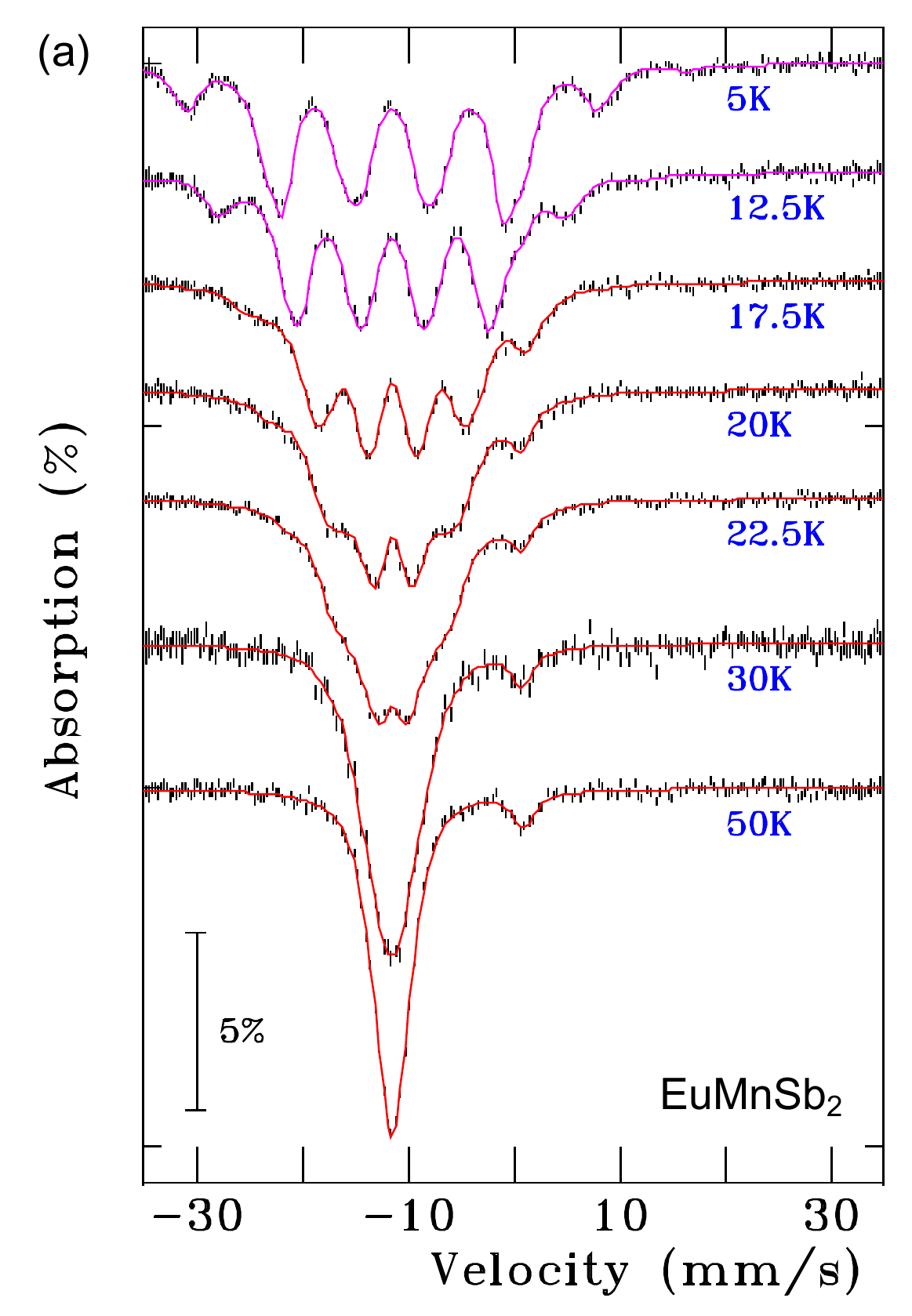}
	\includegraphics[width=1\linewidth]{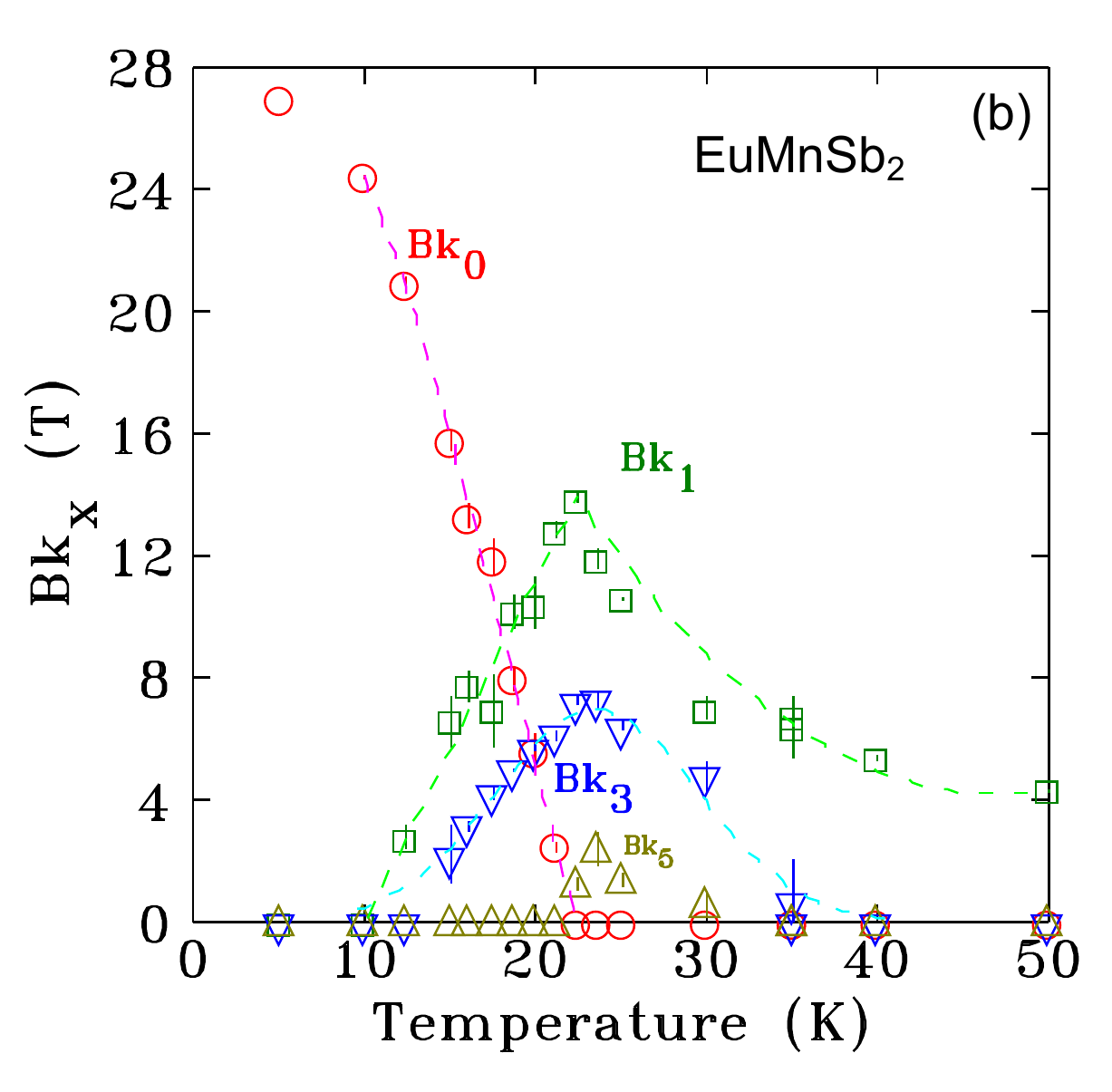}
	\caption{\label{Fig:Moss_Freq} (a) $^{151}$Eu M\"ossbauer spectra of EuMnSb$_2$ at various temperatures. Solid
lines are fits using the two models described in the text: Magenta lines -- full
Hamiltonian, used for $T=5$ and $12.5$~K; red lines -- incommensurate modulation.
(b) Amplitudes of the Fourier components ($Bk_x$) used to fit the modulated
structure, as described is the text.}
\end{figure}

The $^{151}$Eu M\"ossbauer spectrum of EuMnSb$_2$ at $T=5$~K shown in Fig.~\ref{Fig:Moss_Freq}(a) is typical of a magnetically ordered Eu$^{2+}$ compound. There is a sharp, well-split magnetic pattern with a hyperfine field of $B_{\text{hf}}=27.0(1)$~T and an isomer shift of $-11.4(1)$~mm/s. In addition there is a small [$4(1)$\%] contribution from a Eu$^{3+}$ impurity with an isomer shift of $+0.7(1)$~mm/s.  The impurity was either introduced during handling of the sample or is associated with residual flux from the growth process. This component is more apparent at higher temperatures [e.g.\ at $50$~K in Fig.~\ref{Fig:Moss_Freq}(a)]. It is not magnetic and does not affect the fitting. The evolution of the spectra on warming is not typical of a simple AFM~$\rightarrow$~PM transition: above $12.5$~K the lines clearly broaden, and the spectra develop increased weight towards the center of the pattern, reflecting the growing presence of a distribution of hyperfine fields.

We found that the spectra above $T=10$~K are best fitted by assuming a model in which the Eu moments (and, by extension, $B_{\text{hf}}$) develop an incommensurate modulated structure on warming. Following Bonville et al.\ \cite{Bonville_2001} and Maurya et al.\ \cite{Maurya_2014}, we denote the AFM propagation vector as $\mathbf{k}$ in this subsection instead of $\bm{\tau}$, and assume that the modulation in $\bm{\mu}$ along $\mathbf{k}$ can be written in terms of its Fourier components and that $B_{\text{hf}}$ is a linear function of $\mu$ at any given site. Then the variation of $B_{\text{hf}}$ with distance $x$ along $\mathbf{k}$ can be written as \cite{Bonville_2001}
\begin{equation}
	B_{\text{hf}}(kx) = Bk_0 + \sum^n_{l=0} Bk_{2l+1} \sin[(2l+1)kx]\ .
	\label{eqn:fourier}
\end{equation}
$Bk_{n}$ are the odd Fourier coefficients of the field modulation and $kx$ is a position in reciprocal space along the direction of $\mathbf{k}$. As $+B_{\text{hf}}$ and $-B_{\text{hf}}$ are indistinguishable, $kx$ only needs to run over half of the modulation period. Variations of this modeling have been used to fit spectra for EuPdSb \cite{Bonville_2001}, Eu$_4$PdMg \cite{Ryan_2015}, Eu(Co$_{1-x}$Ni$_x$)$_{2-y}$As$_2$ \cite{Sangeetha_2020}, and EuIn$_2$As$_2$ \cite{Riberolles_2021}.

Figure~\ref{Fig:Moss_Freq}(b) shows the evolution of the fitted Fourier components. As the constant term $Bk_0$ falls with increasing $T$, the fundamental term $Bk_1$ and third harmonic term $Bk_3$ develop and rapidly dominate the fits as the distribution of $B_{\text{hf}}$ becomes broader.  By $T\approx22$~K $Bk_0$ is gone, and only the modulated components remain. The broadening of $B_{\text{hf}}$ appears to persist past $250$~K. However, the patterns are not well-enough resolved to assign a meaningful form to the distribution, but the feature associated with Eu$^{2+}$ continues to sharpen on warming.   This residual contribution to $B_{\text{hf}}$ is likely due to short-range magnetic correlations of the Eu.  The neutron diffraction data presented below find evidence for such correlations above \TEuo.

\begin{figure}[]
	\centering
	\includegraphics[width=1\linewidth]{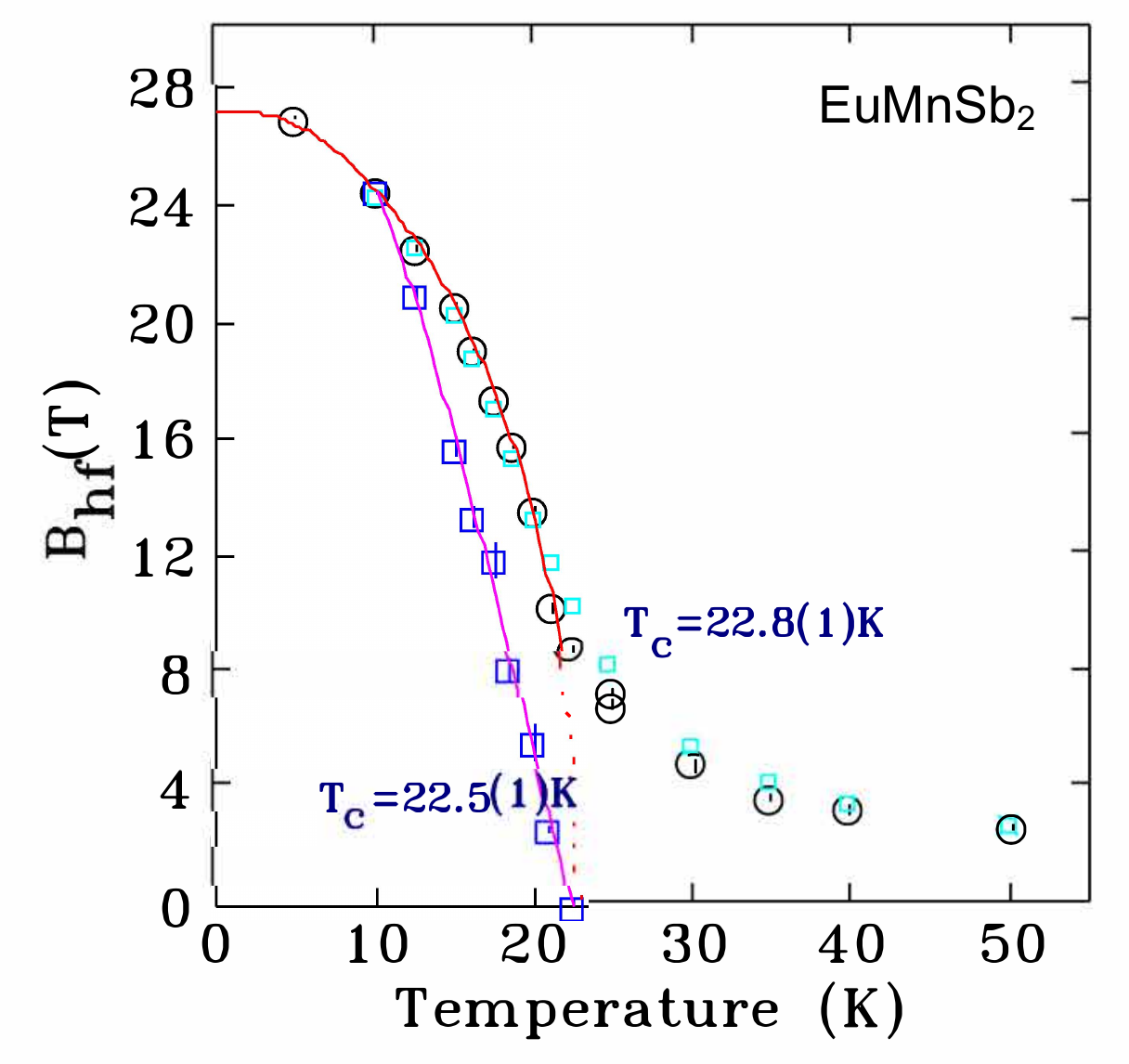}
	\caption{\label{Fig:Moss_OP} Temperature dependence of the hyperfine field for EuMnSb$_2$ derived from the two models used to fit the spectra showing evidence for two transitions. The black circles show $B_{\text{hf}}$ derived from a full Hamiltonian fit and represents an average that does not take account of any distribution. The solid red line is a $J=\frac{7}{2}$ Brillouin function, expected for Eu$^{2+}$ and yielding an extrapolated transition temperature of $\approx22.8(1)$~K. The small cyan squares show the average hyperfine field $\langle B_{\text{hf}} \rangle$ derived from the modulated model. The larger blue squares show the behavior of $Bk_0$, the constant term in the Fourier expansion, which breaks away from $\langle B_{\text{hf}} \rangle$ above $10$~K and reaches zero at $22.5(1)$~K. The residual contributions seen above $25$~K are likely due to short-range magnetic correlations in the Eu sublattice. }
\end{figure}

Figure~\ref{Fig:Moss_OP} shows the temperature dependence of $B_{\text{hf}}$ derived from the two models. The larger black circles show $\langle B_{\text{hf}} \rangle$ derived from the simple one-site full Hamiltonian model showing relatively conventional behaviour (other than the line broadening) up to $T=20$~K that can be fitted to the expected $J=\frac{7}{2}$ Brillouin function for Eu$^{2+}$ yielding an extrapolated magnetic transition temperature of $22.8(1)$~K. The average hyperfine field derived from the modulated fits (smaller cyan squares) tracks $\langle B_{\text{hf}} \rangle$ very well, showing that the two models are consistent. The moment modulation develops between $10$~K and $12.5$~K and $Bk_0$ (shown as the blue squares) falls away from $\langle B_{\text{hf}} \rangle$ quite rapidly. A simple linear fit yields an endpoint of $22.5(1)$~K, consistent with the extrapolation of $\langle B_{\text{hf}} \rangle$.

The $^{151}$Eu M\"ossbauer analysis therefore suggests the presence of two magnetic events associated with the Eu sublattice: initial ordering into an incommensurate structure at $T\approx22$~K that progressively squares up on cooling. The process completes just below $12.5$~K.  On the other hand, as shown below, our single-crystal neutron diffraction data show a commensurate AFM propagation vector below both \TEuo\ and \TEut.   Thus, the incommensurate AFM Eu order between \mbox{\TEut~$\alt T<$~\TEuo} deduced from  M\"ossbauer appears at odds with the commensurate AFM order seen by neutron diffraction. However, M\"ossbauer data for other Eu intermetallics indicate a commensurate AFM ground state and an incommensurate AFM propagation vector that develops as $T$ approaches $T_{\text{N}}$ \cite{Bonville_2001,  Maurya_2014, Sangeetha_2020}. It has been suggested that the development of the incommensurate propagation vector is tied to the presence of short-range magnetic correlations which persist even for $T<T_{\text{N}}$, although they eventually disappear with decreasing $T$ \cite{Sangeetha_2020}.  As we show below, evidence for short-range magnetic correlations is found in our neutron diffraction data for $T\agt$~\TEuo\ which provides consistency between the neutron and M\"ossbauer results.

\subsection{Single-Crystal Neutron Diffraction} \label{SubSec:Neutron}

\begin{figure}[h!]
	\centering
	\includegraphics[width=1\linewidth]{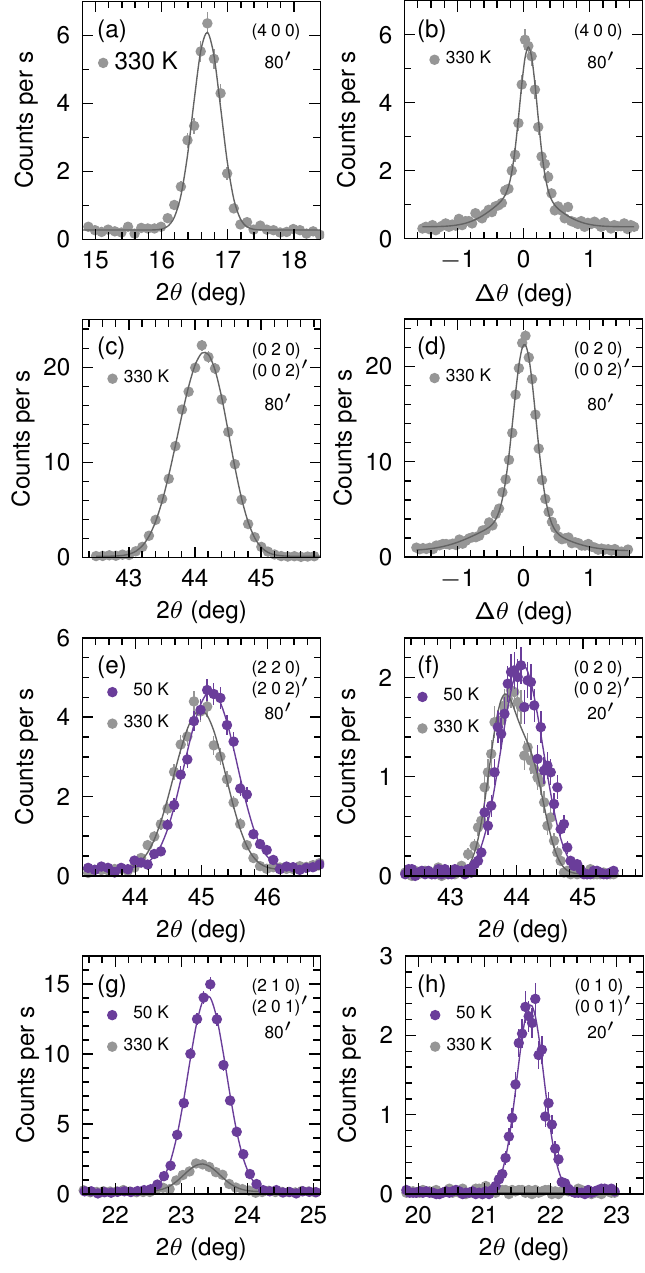}
	\caption{\label{Fig:peaks_high_T} 	Diffraction data for EuMnSb$_2$ from longitudinal ($\theta$-$2\theta$) and  rocking ($\Delta\theta$) scans made at various temperatures using a neutron wavelength of $1.638$~\AA.  The label $80^{\prime}$ ($20^{\prime}$) corresponds to $80^{\prime}$ ($20^{\prime}$) S\"{o}ller-slit collimators being used after the sample.  Lines are fits as discussed in the text.  [(a), (c)] $\theta$-$2\theta$ scan and [(b), (d)] rocking scan data for $(4\,0\,0)$ [(a),(b)] and $(0\,2\,0)/(0\,0\,2)^{\prime}$ [(c),(d)] at $T=330$~K.  [(e), (g)]  $\theta$-$2\theta$ scan data for $(2\,2\,0)/(2\,0\,2)^{\prime}$ (e) and $(2\,1\,0)/(2\,0\,1)^{\prime}$ (g) taken at $330$ and $50$~K.  [(f), (h)] $\theta$-$2\theta$ scan data for $(0\,2\,0)/(0\,0\,2)^{\prime}$ (f) and $(0\,1\,0)/(0\,0\,1)^{\prime}$ (h) at $330$ and $50$~K  using the tighter collimation.  }
\end{figure}

\begin{figure}[]
	\centering
	\includegraphics[width=1.0\linewidth]{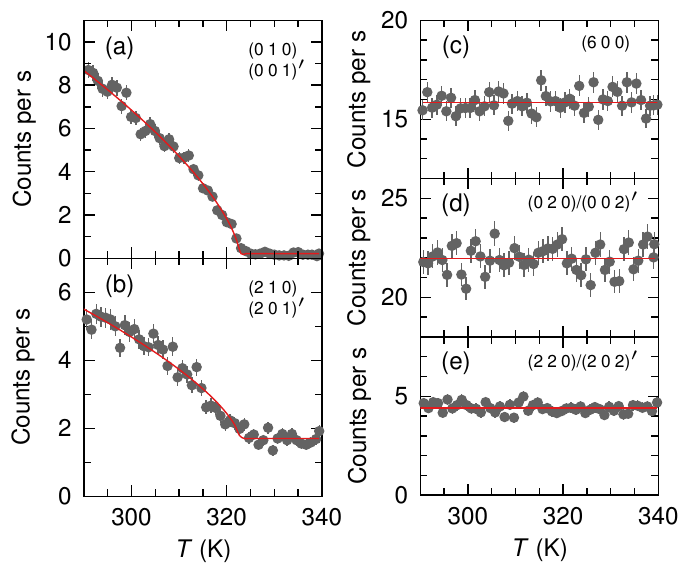}
	\caption{\label{Fig:HighT_OP} 	Heights of the $(0\,1\,0)/(0\,0\,1)^{\prime}$ (a), $(2\,1\,0)/(2\,0\,1)^{\prime}$ (b), $(6\,0\,0)$ (c), $(0\,2\,0)/(0\,0\,2)^{\prime}$ (d), and $(2\,2\,0)/(2\,0\,2)^{\prime}$ (e) Bragg peaks for EuMnSb$_2$ as functions of temperature.  Lines in (a) and (b) are fits to $y_0+(T_{\text{N}}-T)^{2\beta}$ where $y_0$ is a constant, $T_{\text{N}}$ is the N\'{e}el temperature, and $\beta$ is the critical exponent for the magnetic order parameter.  The fits give $\beta\approx0.3$.  Lines in (c)--(e) are guides to the eye.}
\end{figure}

The three ions in EuMnSb$_2$ are all located at the $4c$ Wyckoff position of space group $Pnma$ which does not have any special reflection conditions. Thus, only the general reflection conditions apply for the appearance of structural-Bragg peaks. These are: $(0\,k\,l)$ with $k+l=2n$, $(h\,k\,0)$ with $h=2n$, $(h\,0\,0)$ with $h=2n$, $(0\,k\,0)$ with $k=2n$, and $(0\,0\,l)$ with $l=2n$, where $n$ is an integer. Bragg peaks observed at other positions may be due to magnetic ordering.  Writing $Pnma$ in its  unabbreviated form of $P\frac{2_{1}}{n}\frac{2_{1}}{m}\frac{2_{1}}{a}$ is helpful for understanding the data analysis.  

Refinements to the single-crystal diffraction data taken on TOPAZ confirmed that $P\frac{2_{1}}{n}\frac{2_{1}}{m}\frac{2_{1}}{a}$ correctly describes the chemical-unit cell.  Goodness-of-fit and other parameters determined from the  single-crystal refinements used to simultaneously determine the nuclear and magnetic structures are quoted at the end of Section~\ref{SubSec:Neutron} in Table~\ref{Tab:structure}. Refinements for the nuclear structure allowed the isotropic thermal factor $U_\text{iso}$ and the atomic positions within the unit cell ($x$, $y$, and $z$) to vary.  We were unable to get reliable and sensible results when allowing the occupancy of the sites to vary, and, therefore, assumed the sample to be stoichiometric, as found from EDS.  The values of $U_\text{iso}$, $x$, $y$, $z$, and the twin populations determined at $T=50$~K were used for the refinements made for the $12$ and $5$~K data.

 Figures~\ref{Fig:peaks_high_T}(a) and  \ref{Fig:peaks_high_T}(b) show data from longitudinal ($\theta$-$2\theta$)  and transverse (rocking) scans, respectively, made across the $(4\,0\,0)$ structural-Bragg peak at $T=330$~K with $80^{\prime}$ collimators after the sample. Similar data for $(0\,2\,0)/(0\,0\,2)^{\prime}$ are given in Figs.~\ref{Fig:peaks_high_T}(c) and \ref{Fig:peaks_high_T}(d), where as we explain further below, these data contain contributions from two domains due to twinning.  The Miller indices for the twin domain are labeled by a $\prime$ symbol. For both $(4\,0\,0)$ and $(0\,2\,0)/(0\,0\,2)^{\prime}$, the longitudinal-scan data show single resolution-limited peaks centered at the expected positions which are well fit by a  gaussian lineshape.   The rocking-scan data, on the other hand, require fitting two gaussian peaks to fully account for the lineshape: a sharp resolution-limited peak accounting for the majority of the integrated intensity and a slightly broader, much weaker peak centered at a slightly lower rocking angle.  This mosaic indicates the presence of two grains or for $(0\,2\,0)/(0\,0\,2)^{\prime}$ it could correspond to the tail of the Bragg peak from the twin.

\subsubsection{$\bm{T_{\text{N}_\text{Mn}}>T>T_{\text{N}_\text{Eu1}}}$}

\begin{figure*}[]
	\centering
	\includegraphics[width=1\linewidth]{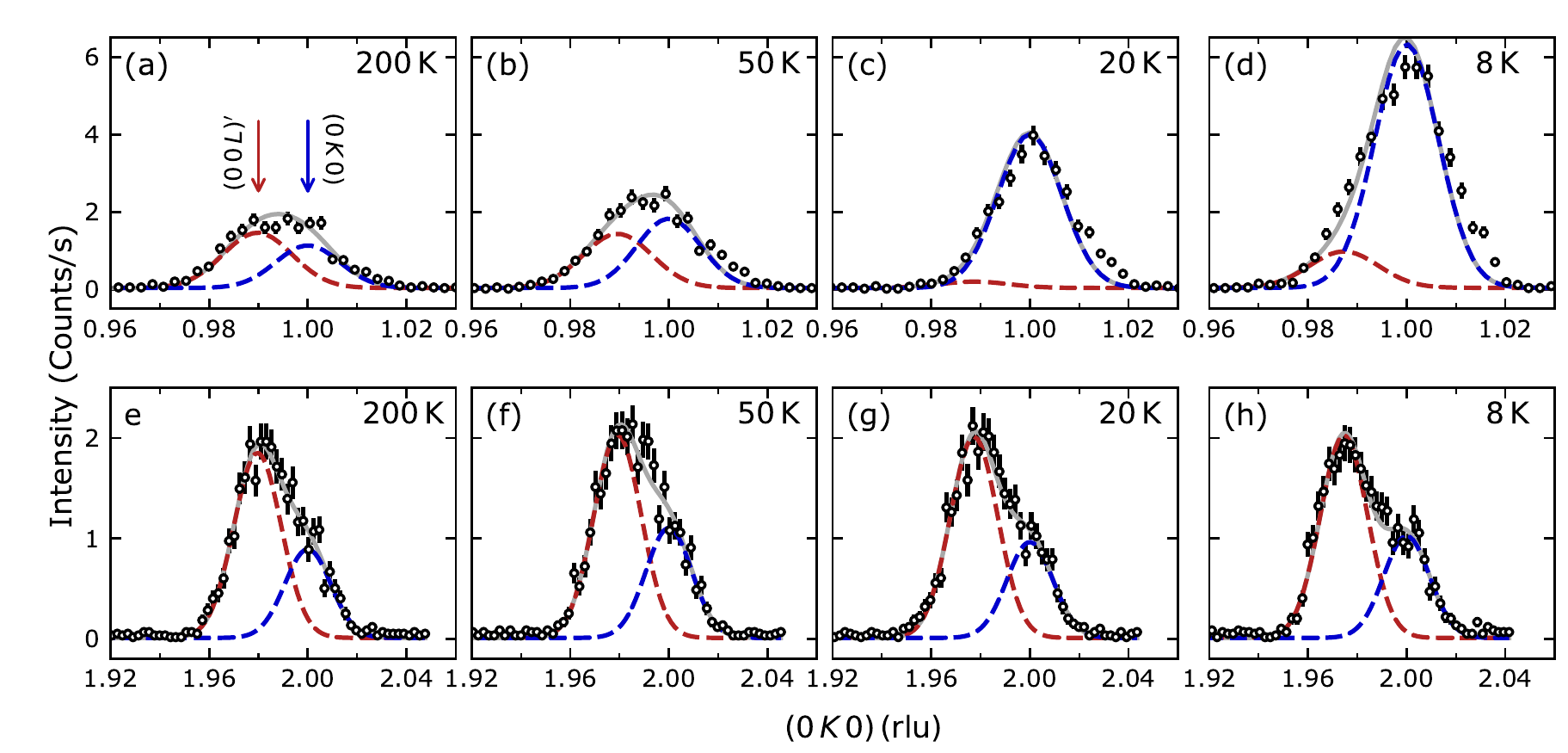}
	\caption{\label{Fig:twin_20min} Diffraction data for EuMnSb$_2$ for $(0\,1\,0)/(0\,0\,1)^{\prime}$ (a-d) and $(0\,2\,0)/(0\,0\,2)^{\prime}$ (e-h) from longitudinal ($\theta$-$2\theta$) scans made at $T=200$~K (a, e), $T=50$~K (b, f), $T=20$~K (c, g) and, $T=8$~K (d, h) using $20^{\prime}$ S\"{o}ller-slit collimators after the sample.  Data are plotted as a function of reciprocal lattice units (rlu) for the twinned directions $(0\,k\,0)$ and $(0\,0\,l)^{\prime}$. Gray lines show the two-gaussian fits to the data  described in the text and red and blue dashed lines show the single-gaussian lineshapes for each twin. The slight shoulder in (b)--(d) above the fitted lineshape shows no substantial temperature dependence and does not affect the analysis presented in the text.}
\end{figure*}

Data in Figs.~\ref{Fig:peaks_high_T}(g), \ref{Fig:peaks_high_T}(h), \ref{Fig:HighT_OP}(a) and \ref{Fig:HighT_OP}(b) demonstrate that magnetic-Bragg peaks emerge upon cooling below \TMn$=323(1)$~K at $(2\,1\,0)/(2\,0\,1)^{\prime}$ and $(0\,1\,0)/(0\,0\,1)^{\prime}$.  The $(2\,1\,0)/(2\,0\,1)^{\prime}$ magnetic-Bragg peak occurs at the same position as a structural-Bragg peak whereas a structural-Bragg peak does not occur at $(0\,1\,0)/(0\,0\,1)^{\prime}$. Magnetic-Bragg peaks at these positions highlight the need to consider twinning of the crystal which can result in misidentifying the Miller indices of a magnetic-Bragg peak.  We were able to resolve both twin domains by inserting $20^{\prime}$ collimators after the sample.  This allowed us to simultaneously resolve $(h\,k\,0)$ and $(h\,0\,l)^{\prime}$ Bragg peaks, but significantly increased the counting time.

Figures~\ref{Fig:twin_20min}(a)--\ref{Fig:twin_20min}(d) show data from longitudinal scans across the  $(0\,1\,0)$ and $(0\,0\,1)^{\prime}$ magnetic-Bragg peak positions at $T=200$, $50$, $20$, and  $8$~K, respectively, using $20^\prime$ collimators after the sample. The recorded lineshapes consists of two peaks, where the solid line is a fit to two gaussian peaks and the dashed lines show the $(0\,1\,0)$ and $(0\,0\,1)^{\prime}$ peaks in blue and red, respectively
.  Figures~\ref{Fig:twin_20min}(e)--\ref{Fig:twin_20min}(h) show similar data for the $(0\,2\,0)$ and $(0\,0\,2)^{\prime}$ structural-Bragg peaks.    Since the $(0\,1\,0)$ and $(0\,0\,1)^{\prime}$ magnetic-Bragg peaks occur at smaller momentum transfers $Q$ than the $(0\,2\,0)$ and $(0\,0\,2)^{\prime}$ structural-Bragg peaks, the individual $(0\,1\,0)$ and $(0\,0\,1)^{\prime}$ magnetic-Bragg peaks are more challenging to resolve and fit.  The fitted two-gaussian lineshapes were found as follows: for each temperature, we determined the positions of the $(0\,1\,0)$ and $(0\,0\,1)^{\prime}$ peaks using the fitted centers for $(0\,2\,0)$ and $(0\,0\,2)^{\prime}$, and the full width at half maximum (FWHM) for each of the two gaussian peaks fit to $(0\,1\,0)/(0\,0\,1)^{\prime}$ was determined at $200$~K and kept fixed for all other temperatures.

The FWHM of each of the red and blue peaks in Fig.~\ref{Fig:twin_20min} agree with the calculated resolutions of $0.031$ and $0.033$ rlu (reciprocal lattice units) for $(0\,1\,0)$ and $(0\,2\,0)$, respectively.  The data and fits in Figs.~\ref{Fig:twin_20min}(a) and \ref{Fig:twin_20min}(b)  indicate that magnetic-Bragg peaks are present at both $(0\,1\,0)$ and $(0\,0\,1)^{\prime}$ below \TMn\ and that they have almost similar integrated intensities at both $T=200$ and $50$~K.  The populations of the twin domains can be estimated from the differences in area of the $(0\,2\,0)$ and $(0\,0\,2)^{\prime}$ Bragg peaks in Figs.~\ref{Fig:twin_20min}(e)--\ref{Fig:twin_20min}(h), however, results presented below from our refinements to the TOPAZ data give a more accurate quantification of the twins' populations.  This is because TOPAZ utilizes detectors that allow for simultaneous measurement of more than one dimension for each peak and because it recorded a greater number of peaks. The data for $20$ and $8$~K are discussed in subsequent subsections.

To summarize Fig.~\ref{Fig:peaks_high_T}, magnetic ordering occurring  at \TMn\ creates magnetic-Bragg peaks at $(h\,k\,0)$ and $(h\,0\,l)^{\prime}$ positions with $h$ even and $k$ or $l$ odd. Figure~\ref{Fig:HighT_OP} shows the detailed temperature dependence of the height of several structural and magnetic-Bragg peaks.  These data were obtained by setting the spectrometer at the center of a peak and measuring while cooling the sample through \TMn.  $80^{\prime}$ collimators were utilized after the sample, and peaks with a nonzero $k$ or $l$ can contain intensity from the twin domain.

As stated above, the increase in intensity for the $(0\,1\,0)/(0\,0\,1)^{\prime}$ and $(2\,1\,0)/(2\,0\,1)^{\prime}$ peaks seen in Figs.~\ref{Fig:HighT_OP}(a) and \ref{Fig:HighT_OP}(b), respectively, upon cooling below \TMn$=323(1)$~K indicates that magnetic-Bragg peaks are present at these locations.  The finite intensity above \TMn\ for $(2\,1\,0)/(2\,0\,1)^{\prime}$ is due to structural-Bragg peaks.  Figures~\ref{Fig:HighT_OP}(c), \ref{Fig:HighT_OP}(d), and \ref{Fig:HighT_OP}(e) show that no discernible changes in intensity occur upon crossing \TMn\ for the $(6\,0\,0)$, $(0\,2\,0)/(0\,0\,2)^{\prime}$, and $(2\,2\,0)/(2\,0\,2)^{\prime}$ positions.  We additionally  did not observe the appearance of any magnetic-Bragg peaks at $(h\,0\,0)$ with $h$ odd while cooling through \TMn.  The magnetic-Bragg peaks can be analyzed using an AFM propagation vector of $\bm{\tau}=(0,0,0)$.  This AFM propagation vector means that the magnetic-unit and chemical-unit cells have the same dimensions.

\begin{table}
	\caption{ \label{Tab:reps_HighT}  Isotropy-magnetic subgroups and  irreducible representations (Irreps) for the space group for EuMnSb$_2$, $P \frac{2_1}{n} \frac{2_1}{m} \frac{2_1}{a}$ ($Pnma$), with Mn and Eu at Wyckoff position $4c$ and an antiferromagnetic propagation vector of $\bm{\tau} = (0\,0\,0)$. Yes and no indicates the presence and absence, respectively, of a magnetic-Bragg peak.  The experimental observations for \TEuo~$<T<$~\TMn\ and \TEut~$<T<$~\TEuo\ are given in the bottom row.} 
	\begin{ruledtabular}
		\begin{tabular}{lccccc}		
		Irrep & Isotropy Magnetic& $(h\,0\,0)$ & $(h\,0\,0)$ &$(0\,1\,0)$ & $(0\,0\,1)$\\
				 & Subgroups& $h$~even & $h$~odd && \\
			\hline
			\noalign{\vskip 1mm}
			$\text{m} \Gamma_{1+}$ & $P \frac{2_1}{n} \frac{2_1}{m} \frac{2_1}{a}$ & no & yes & yes& yes\\
			\noalign{\vskip 1mm}
			$\text{m} \Gamma_{2+}$ & $P \frac{2_1^{\prime}}{n^{\prime}} \frac{2_1^{\prime}}{m^{\prime}} \frac{2_1}{a}$ & yes & no & no & yes\\
			\noalign{\vskip 1mm}
			$\text{m} \Gamma_{3+}$ & $P \frac{2_1}{n} \frac{2_1^{\prime}}{m^{\prime}} \frac{2_1^{\prime}}{a^{\prime}}$ & no & yes & no & no\\
			\noalign{\vskip 1mm}
			$\text{m} \Gamma_{4+}$ & $P \frac{2_1^{\prime}}{n^{\prime}} \frac{2_1}{m} \frac{2_1^{\prime}}{a^{\prime}}$ & yes & no & no & no\\
			\noalign{\vskip 1mm}
			$\text{m} \Gamma_{1-}$ & $P \frac{2_1}{n^{\prime}} \frac{2_1}{m^{\prime}} \frac{2_1}{a^{\prime}}$ & no & yes & yes & yes\\
			\noalign{\vskip 1mm}
			$\text{m} \Gamma_{2-}$ & $P \frac{2_1^{\prime}}{n} \frac{2_1^{\prime}}{m} \frac{2_1}{a^{\prime}}$ & yes & no & no & yes\\
			\noalign{\vskip 1mm}
			$\text{m} \Gamma_{3-}$ & $P \frac{2_1}{n^{\prime}} \frac{2_1^{\prime}}{m} \frac{2_1^{\prime}}{a}$ & no & yes & no & no\\
			\noalign{\vskip 1mm}
			$\text{m} \Gamma_{4-}$ & $P \frac{2_1^{\prime}}{n} \frac{2_1}{m^{\prime}} \frac{2_1^{\prime}}{a}$ & yes & no & yes & no\\
			\noalign{\vskip 2mm}
			\multicolumn{2}{c}{Experiment} & no & yes & yes & yes\\
		\end{tabular}
	\end{ruledtabular}
\end{table}

Consistent with Landau theory, a second-order (continuous) phase transition will decrease the symmetry of a system accordant with a group-subgroup relation.  We used $\textsc{isodistort}$ \cite{isodistort,Stokes_2006} to determine the irreducible representations for magnetic order consistent with space group $P\frac{2_1}{n}\frac{2_1}{m}\frac{2_1}{a}$ and $\bm{\tau}=(0,0,0)$ with magnetic Eu and Mn at the $4c$ Wyckoff position. Table~\ref{Tab:reps_HighT} shows the results, where each irreducible representation corresponds to an isotropy-magnetic subgroup of $P\frac{2_1}{n}\frac{2_1}{m}\frac{2_1}{a}$.    We find that out of the eight irreducible representations both $\text{m} \Gamma_{1+}$ and $\text{m} \Gamma_{1-}$ allow for the existence of magnetic-Bragg peaks at both $(0\,1\,0)$ and $(0\,0\,1)^{\prime}$ as seen in Figs.~\ref{Fig:twin_20min}(a) and \ref{Fig:twin_20min}(b).  $\text{m} \Gamma_{1+}$ is ruled out because it only allows for $\bm{\mu}\parallel\mathbf{b}$ which would necessitate magnetic-Bragg peaks at $(h\,0\,0)$, $h$ odd, positions. This leaves $\text{m}\Gamma_{1-}$ and magnetic space group (MSG) $P \frac{2_1}{n^{\prime}} \frac{2_1}{m^{\prime}} \frac{2_1}{a^{\prime}}$.

\begin{figure*}[]
	\centering
	\includegraphics[width=1.0\linewidth]{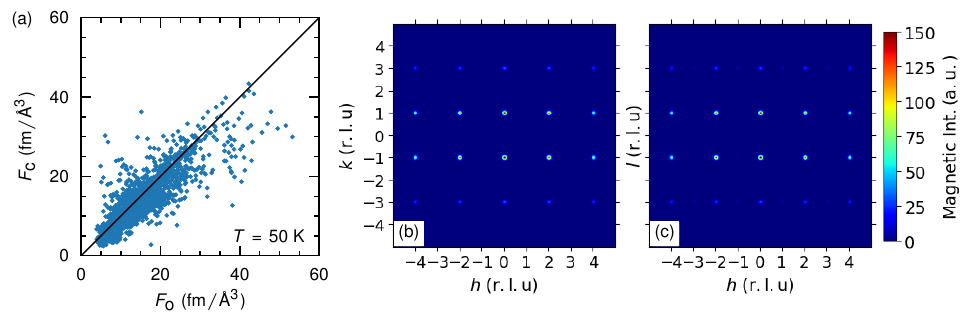}
	\caption{\label{Fig:Jana_50K} (a) The calculated versus observed structure factors for both nuclear and magnetic Bragg peaks from a refinement to $T=50$~K data for EuMnSb$_2$ taken on TOPAZ using the antiferromagnetic structure in Fig.~\ref{Fig:crys_struct}(e).  [(b), (c)] Simulated $(h\,k\,0)$ (b) and $(h\,0\,l)$ (c) diffraction patterns for the $50$~K antiferromagnetic order made with \textsc{mag2pol} \cite{Qureshi_2019}.}
\end{figure*}

Determination of whether the Mn, Eu, or both order below $T_{\text{N}_{\text{Mn}}}$ requires comparing the integrated intensities of the magnetic-Bragg peaks to those calculated for trial magnetic structures. Fortunately, the $4c$ Wyckoff positions occupied by Mn and Eu make the intensity of certain magnetic-Bragg peaks very sensitive to the direction of $\bm{\mu}$ and the positions of the magnetic atoms.  We first discuss some general observations we can make about the possible AFM structure and then present results from a single-crystal refinement that determined the AFM structure.

MSG $P \frac{2_1}{n^{\prime}} \frac{2_1}{m^{\prime}} \frac{2_1}{a^{\prime}}$ allows for  $\bm{\mu}$ to have components lying within the $\mathbf{ac}$ plane.  The existence of the $(0\,1\,0)$ and $(0\,0\,1)^{\prime}$ magnetic-Bragg peaks points to $\bm{\mu}$ having a component along $\mathbf{a}$ whereas the absence of magnetic-Bragg peaks for $(h\,0\,0)$, $h$ odd, means that the component of $\bm{\mu}$ along $\mathbf{c}$ is zero. To consider whether both or either $\bm{\mu}_{\text{Mn}}$ and $\bm{\mu}_{\text{Eu}}$ have a finite component along $\mathbf{a}$, we simulated diffraction patterns using trial structures consistent with $P \frac{2_1}{n^{\prime}} \frac{2_1}{m^{\prime}} \frac{2_1}{a^{\prime}}$ . We found that magnetic-Bragg peaks existing at both $(2\,1\,0)$ and $(2\,0\,1)^{\prime}$ with no magnetic-Bragg peaks at $(2\,2\,0)$ and $(2\,0\,2)^{\prime}$ would indicate a component of $\bm{\mu}_{\text{Mn}}$ lying along $\mathbf{a}$.  On the other hand, no magnetic-Bragg peak at $(2\,1\,0)/(2\,0\,1)^{\prime}$ but a magnetic Bragg peak at $(2\,2\,0)/(2\,0\,2)^{\prime}$ would indicate a component of $\bm{\mu}_{\text{Eu}}$ lying along $\mathbf{a}$.  From Figs.~\ref{Fig:peaks_high_T}(e) and \ref{Fig:peaks_high_T}(g) we see that a magnetic-Bragg peak occurs only at $(2\,1\,0)/(2\,0\,1)^{\prime}$.  Next, if both the Mn and Eu moments were ordered along $\mathbf{a}$ we find that $(0\,1\,0)$ and $(0\,0\,1)^{\prime}$ would have very dissimilar intensities.  Figures~\ref{Fig:twin_20min}(a) and \ref{Fig:twin_20min}(b)  show that this is not the case.  This means that, within our limit of detection, only AFM ordering of the Mn occurs upon cooling through $T_{\text{N}_{\text{Mn}}}$ with $\bm{\mu}_{\text{Mn}}$ lying along $\mathbf{a}$.

The magnetic structure was fully determined by single-crystal refinements using $1944$ structural and $70$ magnetic-Bragg peaks recorded on TOPAZ at $T=50$~K.  The best refinement yielded MSG $P \frac{2_1}{n^{\prime}} \frac{2_1}{m^{\prime}} \frac{2_1}{a^{\prime}}$ with $\mu_{\text{Mn}}=3.0(2)~\mu_{\text{B}}$ lying along $\mathbf{a}$ and $\text{GOF}=6.80$. The AFM order is  C-type, as shown in Fig.~\ref{Fig:crys_struct}(e).  The goodness-of-fit parameter is defined as $\text{GOF}=[\sum{w(F_{\text{o}}-F_{\text{c}})^2}/(m-n)]^{1/2}$ where the sum is over the recorded integrated intensities of the Bragg peaks, $F_{\text{o}}$ is the observed value of the structure factor determined from the integrated intensities, $F_{\text{c}}$ is the calculated value of the structure factor, $m$ is the total number of Bragg peaks used, $n$ is the number of refined parameters, and $w$ is the reciprocal of the sum of a spectrometer specific parameter and the variance associated with each value of $F_{\text{o}}$.  A plot of $F_{\text{c}}$ versus $F_{\text{o}}$ is shown in Fig.~\ref{Fig:Jana_50K}(a) and Figs.~\ref{Fig:Jana_50K}(b) and \ref{Fig:Jana_50K}(c) show simulated diffraction patterns for the AFM structure.  Other than the exact value for $\mu_{\text{Mn}}$, the AFM structure we find for \TEuo$<T<$\TMn\ is the same as those previously reported for both powder and single-crystal samples \cite{Soh_2019, Gong_2020, Zhang_2022}.

The TOPAZ data cover far more of $\mathbf{Q}$ space than our triple-axis measurements and therefore present far more Bragg peaks. Consistent with the triple-axis data in Fig.~\ref{Fig:twin_20min}, the TOPAZ data also indicate that presence of  two twins where the $b$ and $c$ axes are switched between the twins. Our refinement to the TOPAZ data finds that the populations of the two twin domains are $60$\% and $40$\% with an uncertainty of $1$\%.  

\subsubsection{$\bm{T_{\text{N}_\text{Eu1}}>T>T_{\text{N}_\text{Eu2}}}$}

\begin{figure}[]
	\centering
	\includegraphics[width=1.0\linewidth]{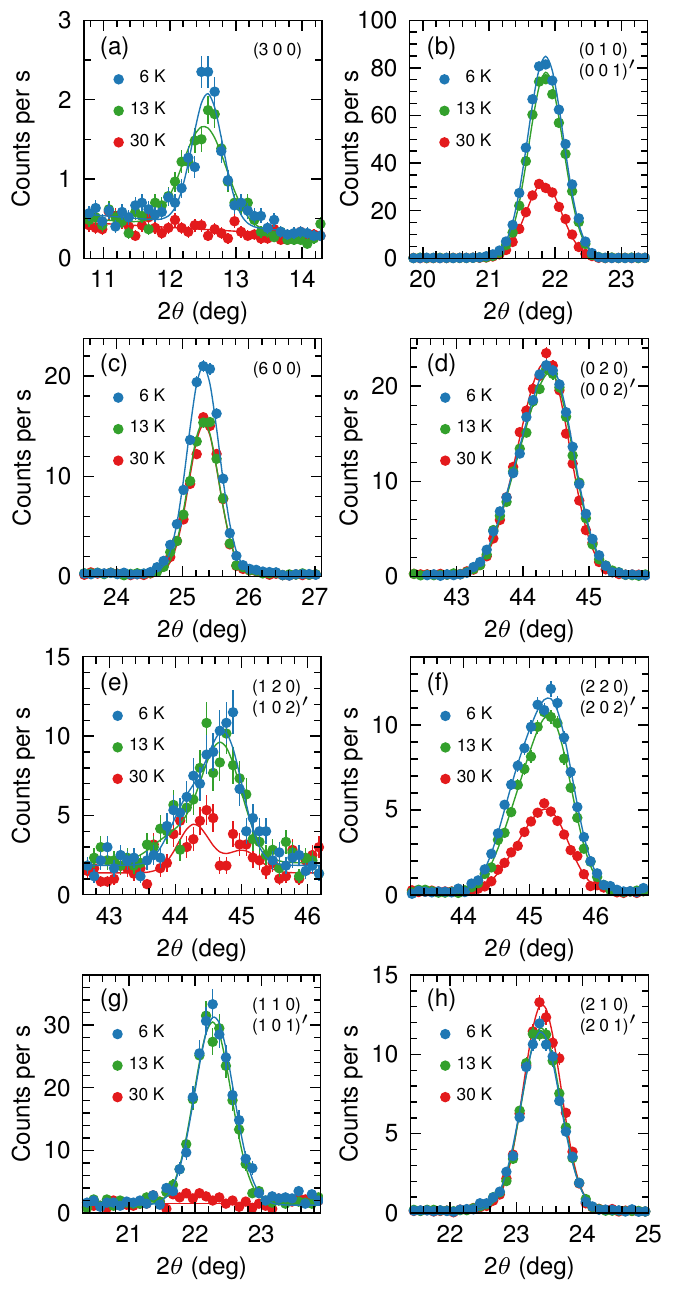}
	\caption{\label{Fig:LowT_peaks} 	Data for EuMnSb$_2$ from longitudinal ($\theta$-$2\theta$) scans across the  $(3\,0\,0)$ (a), $(0\,1\,0)/(0\,0\,1)^{\prime}$ (b), $(6\,0\,0)/(6\,0\,0)^{\prime}$ (c), $(0\,2\,0)/(0\,0\,2)^{\prime}$ (d), $(1\,2\,0)/(1\,0\,2)^{\prime}$ (e), $(2\,2\,0)/(2\,0\,2)^{\prime}$ (f), $(1\,1\,0)/(1\,0\,1)^{\prime}$ (g), and $(2\,1\,0)/(2\,0\,1)^{\prime}$ (h) positions at $T=30$, $13$, and $6$~K. Lines are fits to either the gaussian or two gaussian lineshapes with constant offsets described in the text.}
\end{figure}

\begin{figure}[]
	\centering
	\includegraphics[width=1.0\linewidth]{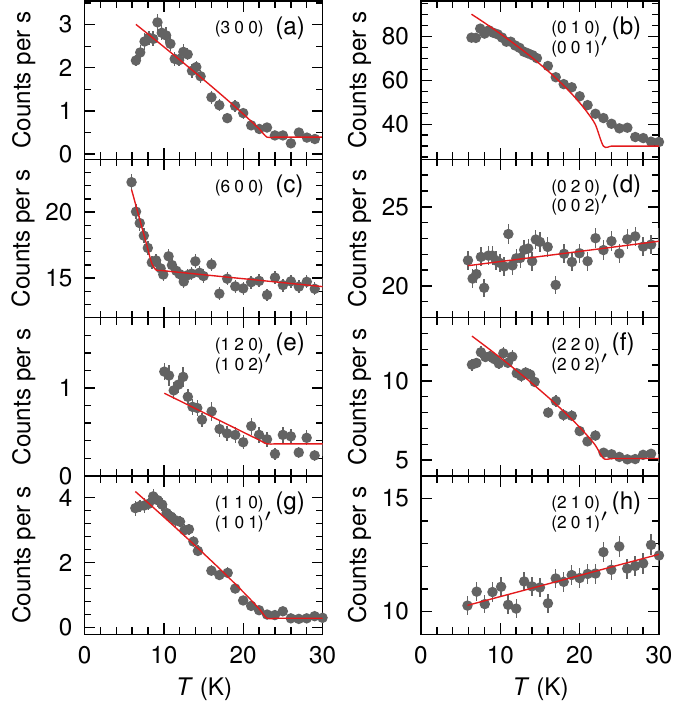}
	\caption{\label{Fig:LowT_OP} 	The intensities of the $(3\,0\,0)$ (a), $(0\,1\,0)/(0\,0\,1)^{\prime}$ (b), $(6\,0\,0)$ (c), $(0\,2\,0)/(0\,0\,2)^{\prime}$ (d), $(1\,2\,0)/(1\,0\,2)^{\prime}$ (e), $(2\,2\,0)/(2\,0\,2)^{\prime}$ (f), $(1\,1\,0)/(1\,0\,1)^{\prime}$ (g), and $(2\,1\,0)/(2\,0\,1)^{\prime}$ (h) Bragg peaks for EuMnSb$_2$ as functions of temperature. A sharp change in intensity below \TEuo and/or \TEut is due to magnetic ordering. Lines are guides to the eye.}
\end{figure}

Figures~\ref{Fig:LowT_peaks} and \ref{Fig:LowT_OP} show that additional magnetic-Bragg peaks emerge below \TEuo\ and\TEut.  They are due to AFM ordering transitions  at   $T_{\text{N}_{\text{Eu1}}}=23(1)$~K and  $T_{\text{N}_{\text{Eu2}}}=9(1)$~K.  We describe our determination of the magnetic order below \TEuo\ in this subsection and our determination of the magnetic order below \TEut\ in the next subsection.

New magnetic-Bragg peaks appear at $(h\,0\,0)$, $h$ odd, positions when cooling below \TEuo, as  seen for $(3\,0\,0)$ in Fig.~\ref{Fig:LowT_peaks}(a) and \ref{Fig:LowT_OP}(a).   Magnetic-Bragg peaks are allowed at these positions for MSG $P \frac{2_1}{n^{\prime}} \frac{2_1}{m^{\prime}} \frac{2_1}{a^{\prime}}$. Similar to the higher temperature data, the lines in Fig.~\ref{Fig:LowT_peaks} show fits to either a gaussian [Figs.~\ref{Fig:LowT_peaks}(a)--\ref{Fig:LowT_peaks}(c), \ref{Fig:LowT_peaks}(g), and \ref{Fig:LowT_peaks}(h)] or two-gaussian lineshape [Figs.~\ref{Fig:LowT_peaks}(d)--\ref{Fig:LowT_peaks}(f)].

Data in Fig.~\ref{Fig:LowT_peaks}(b) show that the $(0\,1\,0)/(0\,0\,1)^{\prime}$  magnetic-Bragg peak continues to grow upon cooling below \TEuo. Data in Fig.~\ref{Fig:twin_20min}(c)  reveal that it is the $(0\,1\,0)$ magnetic-Bragg peak that continues to grow upon cooling below \TEuo\ whereas the $(0\,0\,1)^{\prime}$ peak decreases. Thus, the change in slope of the data in Fig.~\ref{Fig:LowT_OP}(b) is due to an increase in the height of the  $(0\,1\,0)$ magnetic-Bragg peak with decreasing $T$ below \TEuo.  There is no sharp change in intensity associated with crossing \TEuo for either $(6\,0\,0)$ or $(0\,2\,0)/(0\,0\,2)^{\prime}$  as shown in Figs.~\ref{Fig:LowT_peaks}(c) and \ref{Fig:LowT_OP}(c) for $(6\,0\,0)$  and Figs.~\ref{Fig:twin_20min}(f) ,  \ref{Fig:twin_20min}(g), \ref{Fig:LowT_peaks}(d) and \ref{Fig:LowT_OP}(d) for $(0\,2\,0)/(0\,0\,2)^{\prime}$. All of these observations are consistent with the reflection conditions for MSG $P\frac{2_1}{n^{\prime}} \frac{2_1}{m^{\prime}} \frac{2_1}{a^{\prime}}$.

Next, Figs.~\ref{Fig:LowT_peaks}(e) and \ref{Fig:LowT_OP}(e) indicate that a magnetic-Bragg peak appears at $(1\,2\,0)/(1\,0\,2)^{\prime}$ below \TEuo. Weak peaks are also present at $30$~K which indicates that shorter-range magnetic correlations exist at $T>$\TEuo.  As we describe below, the presence of such short-range correlations is consistent with the M\"ossbauer and $C_{\text{p}}$ data.   The existence of a magnetic-Bragg peak at $(1\,2\,0)$ or $(1\,0\,2)^{\prime}$  is also consistent with MSG $P \frac{2_1}{n^{\prime}} \frac{2_1}{m^{\prime}} \frac{2_1}{a^{\prime}}$, as is a magnetic-Bragg peak at $(2\,2\,0)$ or $(2\,0\,2)^{\prime}$.   Data for  $(2\,2\,0)/(2\,0\,2)^{\prime}$ are shown in Figs.~\ref{Fig:LowT_peaks}(f) and \ref{Fig:LowT_OP}(f).
 
 Finally, Figs.~\ref{Fig:LowT_peaks}(g) and \ref{Fig:LowT_OP}(g) show that a magnetic-Bragg peak(s) emerges at $(1\,1\,0)/(1\,0\,1)^{\prime}$ upon cooling  below \TEuo\ whereas data in Figs.~\ref{Fig:LowT_peaks}(h) and \ref{Fig:LowT_OP}(h) show no increase in the magnetic-Bragg peak at $(2\,1\,0)/(2\,0\,1)^{\prime}$. Magnetic-Bragg peaks at $(1\,1\,0)$ and $(1\,0\,1)^{\prime}$ are allowed for MSG $P \frac{2_1}{n^{\prime}} \frac{2_1}{m^{\prime}} \frac{2_1}{a^{\prime}}$. Similar to our above discussion for $(1\,2\,0)/(1\,0\,2)^{\prime}$, the weak peak apparent at $(1\,0\,1)^{\prime}$ in Fig.~\ref{Fig:LowT_peaks}(g) for $30$~K indicates that magnetic correlations begin forming  above \TEuo. 

 Having established that the magnetic-Bragg peaks existing for \TEut$<T<$\TEuo\ are consistent with the reflection conditions for MSG $P \frac{2_1}{n^{\prime}} \frac{2_1}{m^{\prime}} \frac{2_1}{a^{\prime}}$, we next discuss some general considerations for the AFM structure and then present results from single-crystal refinements.  First, for this MSG magnetic-Bragg peaks existing at $(h\,0\,0)$, $h$ odd, means that either or both $\bm{\mu}_{\text{Mn}}$ and $\bm{\mu}_{\text{Eu}}$ have a component along $\mathbf{c}$.    From simulated diffraction patterns, we find that the increase in integrated intensities for the  $(1\,1\,0)$ and $(1\,0\,1)^{\prime}$ magnetic-Bragg peaks is only sensitive to an increase in $\bm{\mu}_{\text{Eu}}$ lying along $\mathbf{c}$. Thus, the increase in height of the $(1\,1\,0)/(1\,0\,1)^{\prime}$ magnetic-Bragg peak when cooling below \TEuo\ which is evident in Figs.~\ref{Fig:LowT_peaks}(g) and \ref{Fig:LowT_OP}(g) is due to the development of a finite component of $\bm{\mu}_{\text{Eu}}$ along $\mathbf{c}$.  This does not, however, strictly rule out that $\bm{\mu}_{\text{Mn}}$ also develops a component along $\mathbf{c}$.
 
 Next, data in Fig.~\ref{Fig:twin_20min}(c) show that the height of the $(0\,1\,0)$ magnetic-Bragg peak greatly increases upon cooling through \TEuo whereas the height of the $(0\,0\,1)^{\prime}$ magnetic-Bragg peak decreases. The simplest explanation for these observations is that $\bm{\mu}_{\text{Eu}}$ develops a finite component parallel to $\mathbf{a}$ as well as a component along $\mathbf{c}$. These observations do not rule out, however, that the component of $\bm{\mu}_{\text{Mn}}$ along $\mathbf{a}$ also increases below \TEuo.
 
 Since both $\bm{\mu}_{\text{Mn}}$ and $\bm{\mu}_{\text{Eu}}$ have finite components along $\mathbf{a}$, the effects due to interference between the Eu and Mn magnetic sublattices on the magnetic structure factor must be taken into account. Thus, one must consider the cases of nearest-neighboring $\bm{\mu_{\text{Mn}}}$ and $\bm{\mu_{\text{Eu}}}$ having $\mathbf{a}$ components aligned either parallel or antiparallel to one another.  This is referred to as the phase between the two magnetic sublattices. Our calculations for the $(0\,1\,0)$ and $(0\,0\,1)^{\prime}$ magnetic-Bragg peak heights demonstrate sensitivity to this phase.  Based on these simulations and the data in Fig.~\ref{Fig:twin_20min}(c) indicating that the height of the $(0\,1\,0)$ magnetic-Bragg peak increases below \TEuo\ whereas the height of the $(0\,0\,1)^{\prime}$ magnetic-Bragg peak decreases, we determined that the $\mathbf{a}$ components of $\bm{\mu}_{\text{Mn}}$ and $\bm{\mu}_{\text{Eu}}$ point antiparallel to each other. 

\begin{figure*}[]
	\centering
	\includegraphics[width=1.0\linewidth]{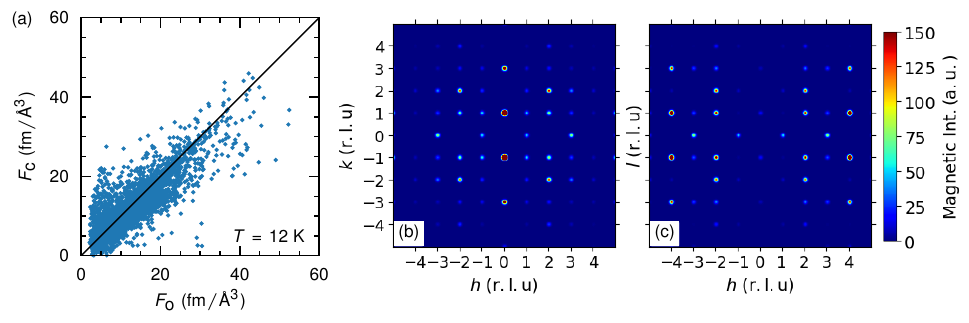}
	\caption{\label{Fig:Jana_12K} (a) The calculated versus observed structure factors for both nuclear and magnetic Bragg peaks from a refinement to $T=12$~K data for EuMnSb$_2$ taken on TOPAZ using the antiferromagnetic structure in Fig.~\ref{Fig:crys_struct}(f).  [(b), (c)] Simulated $(h\,k\,0)$ (b) and $(h\,0\,l)$ (c) diffraction patterns for the $12$~K antiferromagnetic order made with \textsc{mag2pol} \cite{Qureshi_2019}.}
\end{figure*}

We performed single-crystal refinements to our $T=12$~K TOPAZ data using $5330$ nuclear and $374$ magnetic Bragg peaks and various test magnetic structures.   The best refinement gave MSG $P \frac{2_1}{n^{\prime}} \frac{2_1}{m^{\prime}} \frac{2_1}{a^{\prime}}$ and kept the higher temperature Mn sublattice order.   The GOF is $6.11$.   The refinement found $\mu_{\text{Eu}}=4.7 (2)~\mu_{\text{B}}$ with $\bm{\mu_{\text{Eu}}}=4.0 (2)~\mu_{\text{B}}$ along $\mathbf{a}$ and $\bm{\mu_{\text{Eu}}}=2.4 (2)~\mu_{\text{B}}$ along $\mathbf{c}$.  This gives a canting angle of $\phi_{\text{ac}}=31(1)\degree$ away from $\mathbf{a}$ at $12$~K towards $\mathbf{c}$, which is somewhat less than the value of $\phi_{\text{ac}}=41(1)\degree$ found in Ref.~[\onlinecite{Gong_2020}] for $7$~K.  A plot of $F_{\text{c}}$ versus $F_{\text{o}}$ is shown in Fig.~\ref{Fig:Jana_12K}(a) and Figs.~\ref{Fig:Jana_12K}(b) and \ref{Fig:Jana_12K}(c) show simulated diffraction patterns corresponding to the AFM structure.  The AFM structure is illustrated in Fig.~\ref{Fig:crys_struct}(f).  For completeness, a refinement made using the opposite phase between the Mn and Eu magnetic  sublattices resulted in a slightly worse $\text{GOF}$ of $6.14$ with $\bm{\mu}_{\text{Eu}}=3.6 (2)~\mu_{\text{B}}$ along $\mathbf{a}$ and $\bm{\mu}_{\text{Eu}}=2.4 (2)~\mu_{\text{B}}$ along $\mathbf{c}$.

\subsubsection{$\bm{T<T_{\text{N}_\text{Eu2}}}$}
Additional magnetic-Bragg peaks emerge upon cooling below \TEut$=9(1)$~K at $(h\,0\,0)$, $h$ even, positions.  This is shown in Figs.~\ref{Fig:LowT_peaks}(c) and \ref{Fig:LowT_OP}(c) for $(6\,0\,0)$.  Accompanying the appearance of these new magnetic-Bragg peaks is a decrease in the height of the $(3\,0\,0)$ magnetic-Bragg peak, as shown in Fig.~\ref{Fig:LowT_OP}(a).  Figures~\ref{Fig:LowT_OP}(b), \ref{Fig:LowT_OP}(f), and \ref{Fig:LowT_OP}(g) show that the  $(0\,1\,0)/(0\,0\,1)^{\prime}$, $(2\,2\,0)/(2\,0\,2)^{\prime}$, and $(1\,1\,0)/(1\,0\,1)^{\prime}$ magnetic-Bragg peaks either slightly decrease or remain constant upon cooling below \TEut.  Data in Figs.~\ref{Fig:LowT_OP}(d) and \ref{Fig:LowT_OP}(h) indicate that heights of the  $(0\,2\,0)/(0\,0\,2)^\prime$  and $(2\,1\,0)/(2\,0\,1)^\prime$ peaks show no response to cooling through \TEut.

\begin{table}
	\caption{ \label{Tab:reps_LowT}  Maximal-isotropy-magnetic subgroups of  magnetic space group $P \frac{2_1}{n^{\prime}} \frac{2_1}{m^{\prime}} \frac{2_1}{a^{\prime}}$ for EuMnSb$_2$ with $\bm{\tau} = (0\,0\,0)$ and Mn and Eu at the $4c$ Wyckoff position.  The subgroups are given in the setting of $P\frac{2_1}{n}\frac{2_1}{m}\frac{2_1}{a}$.} 
	\begin{ruledtabular}
		\begin{tabular}{c}		
			\noalign{\vskip 0.5mm}
			Isotropy Magnetic Subgroups of $P\frac{2_1}{n^{\prime}}\frac{2_1}{m^{\prime}}\frac{2_1}{a^{\prime}}$\\
			\noalign{\vskip 1mm}
			\hline
			\noalign{\vskip 1mm}
			$P2_1\frac{2_1}{m^{\prime}}\frac{2_1}{a^{\prime}}$\\
			\noalign{\vskip 1mm}
			$P2_12_12_1$\\
			\noalign{\vskip 1mm}
			$P\frac{2_1}{n^{\prime}}11$\\
									\noalign{\vskip 1mm}
			$P1\frac{2_1}{m^{\prime}}1$\\		
									\noalign{\vskip 1mm}
			$P\frac{2_1}{n^{\prime}}\frac{2_1}{m^{\prime}}2_1$\\ 
						\noalign{\vskip 1mm}
			$P~\frac{2_1}{n^{\prime}}2_1\frac{2_1}{a^{\prime}}$\\
						\noalign{\vskip 1mm}
			$P11\frac{2_1}{a^{\prime}}$\\
		\end{tabular}
	\end{ruledtabular}
\end{table}

The appearance of magnetic-Bragg peaks at $(h\,0\,0)$, $h$ even, is inconsistent with MSG $P\frac{2_1}{n^{\prime}} \frac{2_1}{m^{\prime}} \frac{2_1}{a^{\prime}}$. In addition, Table~\ref{Tab:reps_HighT} shows that no MSGs that are maximal-isotropy subgroups of space group $P\frac{2_1}{n} \frac{2_1}{m} \frac{2_1}{a}$ with $\bm{\tau}=(0,0,0)$ and Mn and Eu at the $4c$ Wyckoff position allow for  the simultaneous  appearance of $(h\,0\,0)$, $h$ even, and $(h\,0\,0)$, $h$ odd, magnetic-Bragg peaks. Therefore, it is necessary to consider lower-symmetry MSGs.  To this end, Table~\ref{Tab:reps_LowT} shows the maximal-isotropy subgroups for MSG  $P\frac{2_1}{n^{\prime}} \frac{2_1}{m^{\prime}} \frac{2_1}{a^{\prime}}$ with $\bm{\tau}=(0,0,0)$ and  Mn and Eu at the $4c$ Wyckoff position.

We can immediately exclude MSGs $P2_1\frac{2_1}{m^{\prime}}\frac{2_1}{a^{\prime}}$, $P2_12_12_1$, and $P\frac{2_1}{n^{\prime}}$11 since they forbid magnetic-Bragg peaks at $(h\,0\,0)$, $h$ even.  We then tested our data against calculations for various trial magnetic-moment configurations under MSG $P1\frac{2_1}{m^{\prime}}1$, which corresponds to the MSG found in Ref.~[\onlinecite{Soh_2019}] for the low-temperature AFM structure but in a different crystallographic setting.  We found that $\bm{\mu}_{\text{Mn}}$ would need to develop a significant FM component along $\mathbf{c}$ to produce a magnetic-Bragg peak at $(3\,0\,0)$ with the height observed in Fig.~\ref{Fig:LowT_peaks}(a).  This would also require much weaker magnetic-Bragg peaks at  $(1\,1\,0)/(1\,0\,1)^{\prime}$ than we observe in Fig.~\ref{Fig:LowT_peaks}(g): the intensity of $(1\,1\,0) $ would be zero and the intensity of $(1\,0\,1)^{\prime}$ would be quite weak.  Thus, we found it unlikely that MSG $P1\frac{2_1}{m^{\prime}}1$ describes the AFM order below \TEut\ for our sample.
 
 \begin{figure*}[]
 	\centering
 	\includegraphics[width=1.0\linewidth]{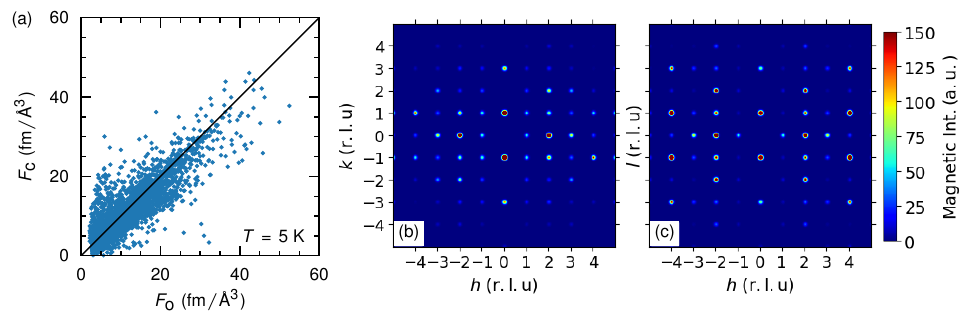}
 	\caption{\label{Fig:Jana_5K} (a) The calculated versus observed structure factors for both nuclear and magnetic Bragg peaks from a refinement to $T=5$~K data for EuMnSb$_2$ taken on TOPAZ using the antiferromagnetic structure in Fig.~\ref{Fig:crys_struct}(g).  [(b), (c)] Simulated $(h\,k\,0)$ (b) and $(h\,0\,l)$ (c) diffraction patterns for the $5$~K antiferromagnetic order made with \textsc{mag2pol} \cite{Qureshi_2019}.}
 \end{figure*}
 
 We next made single-crystal refinements using the MSGs in Table~\ref{Tab:reps_LowT} and trial ordered-moment configurations in order to determine the AFM structure at $T=5~K$.   The refinements used $4573$ nuclear and $321$ magnetic-Bragg peaks recorded on TOPAZ.  The results indicated that MSG $P11\frac{2_1}{a^{\prime}}$ with lattice angles $\alpha=\beta=\gamma=90\degree$ best describes the AFM structure with the higher temperature AFM order of the Mn magnetic sublattice maintained.  The ratios of the four domains present for the MSG were set to $30$\%:$30$\%:$20$\%:$20$\% which is in line with the twin populations found at $50$~K.
 
 The best refinement found that $\bm{\mu}_{\text{Eu}}$ develops a component along $\mathbf{b}$ below \TEut. The  components of $\bm{\mu}_{\text{Eu}}$ are found to be  $2.4(4)$, $4.3(2)$, and $2.6(2)~\mu_{\text{B}}$ lying along $\mathbf{a}$, $\mathbf{b}$, and $\mathbf{c}$, respectively. This gives at total Eu-ordered-magnetic moment of $\mu_{\text{Eu}}=5.6(4)~\mu_{\text{B}}$. The $\text{GOF}$ parameter is $6.19$, which is similar to the value found for the higher temperature AFM phases.  The magnetic structure is shown in Fig.~\ref{Fig:crys_struct}(g) and Fig.~\ref{Fig:Jana_5K}(a) shows the results of the refinement.  Figures~\ref{Fig:Jana_5K}(b) and \ref{Fig:Jana_5K}(c) show simulated diffraction patterns corresponding to the AFM structure.
 
 To test other possible solutions, a refinement made with the components of $\bm{\mu}_{\text{Eu}}$  along $\mathbf{a}$ and $\mathbf{c}$ fixed to those found at $T=12$~K [$4.0 (2)~\mu_{\text{B}}$ and $2.4 (2)~\mu_{\text{B}}$ along $\mathbf{a}$ and $\mathbf{c}$, respectively] results in a worse $\text{GOF}$ of $6.22$ and a component of $\bm{\mu}_{\text{Eu}}$ along $\mathbf{b}$ of  $3.5(2)~\mu_{\text{B}}$.  The magnitude of $\bm{\mu}_{\text{Eu}}$ found from both refinements is similar, being $\mu_{\text{Eu}}=5.6(4)~\mu_{\text{B}}$ for the former and $5.8(2)~\mu_{\text{B}}$ for the latter.
 
 As for the other two MSGs not yet ruled out, a refinement using MSG $P\frac{2_1}{n^{\prime}}\frac{2_1}{m^{\prime}}2_1$ yields a similar GOF but does not give a value of the $R$ factor for the magnetic part of the refinement that is lower than that we obtain for our best refinement for MSG $P11\frac{2_1}{a^{\prime}}$.  $R$ is defined as  $R=\sum_i|y_i(obs)-y_i(calc)|/y_i(obs)$ where $y_i(obs)$ and $y_i(calc)$ correspond to the observed and calculated integrated intensities for the measured peak $i$. For MSG $P~\frac{2_1}{n^{\prime}}2_1\frac{2_1}{a^{\prime}}$, the best refinement gives similar values for $R$ and $\text{GOF}$ as our best refinement using MSG $P11\frac{2_1}{a^{\prime}}$, but it indicates a net $\mathbf{M}$ along $\mathbf{b}$ with a component of  $\bm{\mu}_{\text{Eu}}$ along $\mathbf{b}$ of $1.5(3)~\mu_{\text{B}}$.  However, the absence of a clear increase in the intensity of the $(0\,0\,2)^\prime$ peak in Figs.~\ref{Fig:twin_20min}(h), \ref{Fig:LowT_peaks}(d), and \ref{Fig:LowT_OP}(d) upon cooling below \TEut\ tends to rule out this MSG. 
 
 Table~\ref{Tab:structure} summarizes the  parameters in each AFM phase determined from our best refinements.   Lattice parameters for various temperatures are also included.  The value of $\mu_{\text{Eu}}=5.6(4)~\mu_{\text{B}}$ we find for $T=5$~K is lower than the full value of $7~\mu_\text{B}$ expected for Eu$^{2+}$.  $\mu_{\text{Eu}}$ may further increase upon cooling below $5$~K, however, an increase of  $1.2~\mu_\text{B}$  between $5$ and $0$~K is not expected by the curves in Fig.~\ref{Fig:LowT_OP}.  On the other hand, $M(\mu_0H)$ data for EuMnBi$_2$ taken up to $\mu_0H=52$~T find a saturated moment at $\mu_0H=22$~T and $T=1.4$~K of only $6.4~\mu_{\text{B}}$  \cite{Masuda_2016}. This suggests $\mu_{\text{Eu}}<7~\mu_{\text{B}}$ for this structurally related compound.  Similar high-field measurements to determine the saturated moment for EuMnSb$_2$ should be insightful.
 	
 We must also consider that the significant neutron absorption due to  $^{151}$Eu negatively affects the accuracy and precision with which we can determine $\mu_{\text{Eu}}$ and $\mu_{\text{Mn}}$ as well as other parameters.   Lower than expected values for $\mu_{\text{Eu}}$ as determined by  neutron diffraction have been reported for other Eu-containing compounds \cite{CHATTOPADHYAY_1982}, including previous studies on EuMnSb$_2$ which found that $\mu_{\text{Eu}}=5.9(8)~\mu_{\text{B}}$ at $7$~K  \cite{Gong_2020} and $\mu_{\text{Eu}}=5.2(4)~\mu_{\text{B}}$ at $5$~K \cite{Zhang_2022}.  For our measurements, we used standard algorithms to mitigate and correct for the absorption effects.  However, while good, these algorithms are imperfect and make determining the exact uncertainty associated with certain parameters, such as $\mu_{\text{Eu}}$ and $\mu_{\text{Mn}}$, problematic. The uncertainty values for $U_{\text{iso}}$, $x$, $y$, $z$, $\mu_{\text{Mn}}$, $\mu_{\text{Eu}}$, and the components of  $\bm{\mu_{\text{Eu}}}$ quoted in Table~\ref{Tab:structure} correspond to one standard deviation from their refined values.  The actual uncertainty is likely larger and difficult to precisely quantify.
 
 Finally, regarding the M\"ossbauer data presented in Section~\ref{SubSec:Moss} and the presence of short-range magnetic correlations, such correlations would give rise to weak and broad magnetic diffraction peaks.  These peaks would be difficult to resolve below \TEuo\ in the midst of AFM Bragg peaks via typical neutron diffraction measurements.  The presence of thermal-neutron absorbing Eu in EuMnSb$_2$ would further diminish the ability to detect such weak peaks.  Nevertheless, as noted above, there is some evidence in our diffraction data [e.g.\ the shape of the order-parameter curve upon crossing \TEuo\ in Fig.~\ref{Fig:LowT_OP}(b)] for short-range magnetic correlations existing above \TEuo.  The $C_{\text{p}}(T)$ data in Fig.~\ref{Fig:Cp} exhibit a broad peak at \TEuo\ which is also consistent with the presence of short-range magnetic correlations, although subtle structural disorder not resolved in our neutron diffraction, x-ray diffraction, and EDS measurements could also broaden the peak.
 
  \begin{table}
 	\caption{ \label{Tab:structure}  Structural and magnetic parameters for orthorhombic EuMnSb$_2$ at various temperatures.  All of the atoms occupy the $4c$ Wykoff position of space group $Pnma$ ($P\frac{2_1}{n}\frac{2_1}{m}\frac{2_1}{a}$).  The parameters are from refinements made to data taken on TOPAZ.   $\mu_{\text{Eu-a}}$, $\mu_{\text{Eu-b}}$, and $\mu_{\text{Eu-c}}$ are the components of the ordered Eu magnetic moment $\bm{\mu_{\text{Eu}}}$ along $\mathbf{a}$, $\mathbf{b}$, and $\mathbf{c}$, respectively.  GOF is the goodness-of-fit parameter for the combined nuclear and magnetic refinements and is defined in the text. MSG stands for magnetic space group.  The best refinements found that the ordered Mn magnetic moment $\bm{\mu_{\text{Mn}}}$ lies along $\mathbf{a}$ and that the antiferromagnetic order of the Mn sublattice found at $50$~K remains unchanged for $12$ and $5$~K.  The structural parameters $U_{\text{iso}}$, $x$, $y$, and $z$ determined at $50$~K were also used for $12$ and $5$~K, as well as the refined values for the twin populations of $60(1)$\% and $40(1)$\%. $x$, $y$, and $z$ are given in the basis of space group $P\frac{2_1}{n}\frac{2_1}{m}\frac{2_1}{a}$.} 
 	\begin{ruledtabular}
 		\begin{tabular}{cd{2.6}d{2.6}d{2.6}}
 		&\multicolumn{1}{c}{$T=50$~K}&\multicolumn{1}{c}{$12$~K}&\multicolumn{1}{c}{$5$~K}\\
 		\hline\noalign{\vskip 1mm}
 		$a$~(\AA)&22.4958(1)&22.4928(2)&22.4927(2)\\
 		$b$~(\AA)&4.3758(1)&4.3787(1)&4.3773(1)\\
 		$c$~(\AA)&4.3908(1)&4.3791(1)&4.3802(1)\\
 		$V$~(\AA$^3$)&431.45(1)&431.29(1)&431.26(1)\\
 		Eu $U_{\text{iso}}$&0.0066(3)&0.0066&0.0066\\
 		Eu $x$&0.38598(6)&0.38598&0.38598\\
 		Eu $y$&0.25&0.25&0.25\\
 		Eu $z$&0.7712(9)&0.7712&0.7712\\
 		Mn $U_{\text{iso}}$&0.0070(4)&0.0070&0.0070\\
 		Mn $x$&0.2501(3)&0.2501&0.2501\\
 		Mn $y$&0.25&0.25&0.25\\
 		Mn $z$&0.271(2)&0.271&0.271\\
 		Sb $1$ $U_{\text{iso}}$&0.0044(3)&0.0044&0.0044\\
 		Sb $1$ $x$&0.32545(6)&0.32545&0.32545\\
 		Sb $1$ $y$&0.75&0.75&0.75\\
 		Sb $1$ $z$&0.2727(9)&0.2727&0.2727\\
 		Sb $2$ $U_{\text{iso}}$&0.0044(3)&0.0044&0.0044\\
 		Sb $2$ $x$&0.4984(1)&0.4984&0.4984\\
 		Sb $2$ $y$&0.25&0.25&0.25\\
 		Sb $2$ $z$&0.2813(9)&0.2813&0.2813\\
 		$\mu_{\text{Mn}}$~($\mu_{\text{B}}$)&3.0(2)&3.0&3.0\\
 		$\mu_{\text{Eu}}$~($\mu_{\text{B}}$)&0&4.7(2)&5.6(4)\\
 	$\mu_{\text{Eu-a}}$~($\mu_{\text{B}}$)&0&4.0(2)&2.4(4)\\
 	$\mu_{\text{Eu-b}}$~($\mu_{\text{B}}$)&0&0&4.3(2)\\
 	$\mu_{\text{Eu-c}}$~($\mu_{\text{B}}$)&0&2.4(1)&2.6(1)\\
 		\text{GOF}&6.80&6.11&6.19\\
 		\multicolumn{1}{c}{MSG}&\multicolumn{1}{c}{$P\frac{2_1}{n^\prime}\frac{2_1}{m^\prime}\frac{2_1}{a^\prime}$}&\multicolumn{1}{c}{$P\frac{2_1}{n^\prime}\frac{2_1}{m^\prime}\frac{2_1}{a^\prime}$}&\multicolumn{1}{c}{$P11\frac{2_1}{a^\prime}$}\\
 	\end{tabular}
 \end{ruledtabular}
\end{table}
 
\section{Discussion}
 Based on our results, we can make some general conclusions about the microscopic magnetic interactions in the compound.  We start by using the determined AFM structures to make some assumptions about a minimal Heisenberg model for the compound that includes single-ion magnetic anisotropy.  This model has the generic form
\begin{equation}
	\mathcal{H}=\sum_{i,j}\tilde{S}_{i}J_{ij}S_{j}+\sum_{i}\tilde{S}_{i}D_{i}S_{i}\ ,%
\end{equation}
where $J$ is the magnetic exchange, $D$ is magnetic anisotropy, $i$ and $j$ label magnetic ions, and $S$ is spin.  The AFM order of the Mn planes above \TEuo suggests AFM intralayer exchange for NN Mn spins within a plane,  FM interlayer exchange between NN Mn layers, and Mn easy-axis anisotropy preferring the $\mathbf{a}$ direction.  These considerations hold for all temperatures.

The AFM order of the Eu sublattice below both \TEuo\ and \TEut\ points to dominant FM NN intralayer exchange. The NN interlayer exchange between Eu layers depends on whether a Sb$1$-Mn-Sb$1$ block or a Sb$2$ layer separates the Eu layers.  For both cases, the $\mathbf{a}$ and $\mathbf{b}$ components of the Eu interlayer exchange are AFM.  However, the $\mathbf{c}$ component  of the Eu interlayer exchange is FM across Sb$1$-Mn-Sb$1$ blocks and AFM across Sb$2$ layers.  The appearance of a finite component of $\bm{\mu_{\text{Eu}}}$ along $\mathbf{b}$ below \TEut\ rather than \TEuo suggests that the $\mathbf{b}$ component of the interlayer Eu exchange is smaller than the other exchange terms.  It is also possible that the Eu anisotropy changes with decreasing temperature such that the tendency for $\bm{\mu}_{\text{Eu}}$ to lie mostly along $\mathbf{a}$ weakens with decreasing $T$, especially upon crossing below \TEut.

The NN interlayer exchange between Eu and Mn  may be weak, since based on the diffraction results the Eu ordering does not appear to affect the Mn sublattice order.  However, as discussed below, this would differ with a reported result for the related compound EuMnBi$_2$ which finds substantial Eu-Mn coupling based on $M(H)$ data \cite{May_2014}.  Inelastic neutron scattering measurements to determine the magnetic interactions from fits to the spin-wave spectrum are a typical way to sort out the magnetic Hamiltonian.  However, the thermal-neutron absorbing properties of $^{151}$Eu and its $\approx48$\% natural abundance will make planning and performing such measurements especially challenging.

Considering the different AFM structures found below \TEuo\ by powder neutron diffraction \cite{Soh_2019} and the report of only two AFM phases down to $T=7$~K from previous single-crystal neutron diffraction data \cite{Gong_2020}, the finding of a third AFM phase in our samples begs questions concerning the sensitivity of the magnetism to stoichiometry or other aspects of the chemical structure.   Indeed, Ref.~\onlinecite{Zhang_2022} presents a study of single crystals for the series Eu$_{1-x}$Sr$_{x}$MnSb$_{2}$ and  finds a tetragonal chemical unit cell [space group $P\frac{4}{n}mm$ with $a=4.343(6)$~\AA\ and $c=11.17(1)$~\AA\ at $5$~K] for $x=0$.  A $5\%$ Mn deficiency is  reported for the tetrgaonal parent compound (EuMn$_{0.95}$Sb$_2$), and the orthorhombic chemical unit cell is found for $x=0.2$ \cite{Zhang_2022}.  The results suggest sensitivity of both the chemical-unit cell and the AFM order to stoichiometry.

The structures of the AFM order above \TEut\ for our sample are consistent with those previously found for other single-crystal orthorhombic EuMnSb$_2$ samples \cite{Gong_2020}.  Other than the phase between the magnetic sublattices, our $T>$~\TEut\ AFM structures are also consistent with those for  orthorhombic Eu$_{0.8}$Sr$_{0.2}$MnSb$_2$ \cite{Zhang_2022}.  A third AFM transition is also seen in Eu$_{0.8}$Sr$_{0.2}$MnSb$_2$ below $7$~K.  However, it is associated with a change in the ordering of the $\mathbf{c}$ component of $\bm{\mu}_{\text{Eu}}$ from $\rightarrow\rightarrow\leftarrow\leftarrow$ to $\rightarrow\leftarrow\rightarrow\leftarrow$, which is different than the AFM order we find in our EuMnSb$_2$ sample below \TEut.  Notably, with increasing $x$ for Eu$_{1-x}$Sr$_{x}$MnSb$_{2}$, $\bm{\mu}_{\text{Eu}}$ rotates towards $\mathbf{a}$ with $\phi_{\text{ac}}=0\degree$ for $x=0.8$, and a third (lower temperature) phase transition is not observed for $x=0.5$ and $0.8$ \cite{Zhang_2022}. A component of $\bm{\mu_{\text{Eu}}}$ lying along $\mathbf{b}$ is not reported for the series.  To the best of our knowledge, our findings are the first report of a finite component of $\bm{\mu_{\text{Eu}}}$ along $\mathbf{b}$ in EuMnSb$_2$.  To explore the tunability of the magnetism, future work should determine what parameters create the existence of a $\mathbf{b}$ component for $\bm{\mu_{\text{Eu}}}$.

Next, we once again consider EuMnBi$_2$, which has a tetragonal unit cell with space group $I\frac{4}{m}mm$  \cite{May_2014} and hosts Dirac fermions within its $X2$ ($X=$~Bi) layers \cite{Masuda_2016}.  This compound exhibits AFM ordering of the Mn and Eu moments below $\approx315$ and $22$~K, respectively, with both $\bm{\mu}_{\text{Mn}}$ and $\bm{\mu}_{\text{Eu}}$ lying along $\mathbf{c}$ \cite{May_2014}.   The Mn sublattice consists of AFM alignment of NN Mn within each Mn layer but AFM alignment between Mn layers instead of the FM interlayer alignment we observe in EuMnSb$_2$. The AFM structure of the Eu sublattice has an $\uparrow\uparrow\downarrow\downarrow$ stacking of FM Eu layers along $\mathbf{c}$ \cite{Masuda_2018} which also differs from the $\downarrow\uparrow\downarrow\uparrow$ configuration of the $\mathbf{a}$ component of $\bm{\mu}_{\text{Eu}}$ we observe for EuMnSb$_2$.  The AFM structures are similar, however,  in that the Eu layers sandwiching the $X2$ layers are AFM aligned. The dominant AFM intralayer Mn exchange and FM intralayer Eu exchange suggests that an orthorhombic distortion of the tetragonal lattice does not change the sign of the intralayer Mn and Eu exchange since the same is found for EuMnSb$_2$.


 Below $T=22$~K a spin-flop transition is seen in $M(H)$ for EuMnBi$_2$ at $\mu_0H\approx5.4$~T for $\mathbf{H}\parallel\mathbf{c}$ which reorients $\bm{\mu}_{\text{Eu}}$ into the $\mathbf{ab}$ plane \cite{May_2014, Masuda_2020}.  $\bm{\mu}_{\text{Eu}}$ becomes fully polarized along $\mathbf{H}$ above $\approx22$~T at $\approx2$~K where $M$ is taken to be close to the expected fully saturated value of $7~\mu_{\text{B}}$ even though it only reaches $6.4~\mu_{\text{B}}$ \cite{Masuda_2016}.  The absence of a clear signature for a spin-flop transition at $1.8$~K in our data for EuMnSb$_2$ [Fig.~\ref{Fig:Mag}(b)] can be explained by $\bm{\mu}_{\text{Eu}}$ already having canted away from $\mathbf{a}$.
 
 The interlayer-longitudinal resistivity $\rho_{\text{zz}}$ for EuMnBi$_2$ is seen to greatly depend on the magnetic order of the Eu sublattice, as $\rho_{\text{zz}}$ is almost independent of field above $22$~K, and below $22$~K a large enhancement in $\rho_{\text{zz}}$ within the spin-flop region of the phase diagram is seen along with the half-integer-quantum-Hall effect \cite{Masuda_2016, Masuda_2018}.  The exact microscopic mechanism  behind the enhancement  is not clear, however, as well as the exact role of the Mn magnetic sublattice \cite{Masuda_2016, Masuda_2020}.  The transverse magnetoresistance $\rho_{\text{xx}}$ is found to be  large and positive \cite{ Masuda_2016}.  It is linear in $H$ at $50$~K, and quantum oscillations are superimposed on the linear magnetoresistance with a large enhancement of the height of the oscillations occurring at magnetic fields in the spin-flop region for $1.4$~K \cite{ Masuda_2016}.
 
For EuMnSb$_2$, interlayer-longitudinal-resistivity data reported in Refs.~[\onlinecite{Soh_2019}] and [\onlinecite{Gong_2020}] for single-crystal samples indicate a loss of spin-disorder scattering upon cooling below \TEuo\ at $\mu_0H=0$~T. Reference~[\onlinecite{Soh_2019}] also reports an $\approx300$\% increase in resistance with increasing field with a maximum at $\approx5$~T at $T=2$~K, with negative magnetoresistance occurring for higher values of $\mu_0H$.  Angular dependent measurements made at $2$~K show that as the field is increased the interlayer-longitudinal resistivity exhibits growing anisotropy, with clear peaks when the angle between $\mathbf{a}$ and $\mathbf{b}$ is $\phi_{\text{ab}}=0\degree$ and $180\degree$ for $\mu_0H\ge1$~T \cite{Soh_2019}.  This suggests an increase in interlayer scattering for $\mathbf{H}\parallel\mathbf{a}$. However, similar to our discussion for EuMnBi$_2$, the microscopic contributions of the magnetic order to the increase in scattering need to be further elucidated.  Unlike EuMnBi$_2$, quantum oscillations for EuMnSb$_2$ have not been seen for either $M$ data taken down to $1.8$~K and in fields up to $7$~T or for $\rho_{\text{xx}}$ data taken down to $2$~K and in fields up to $13$~T \cite{Soh_2019}. This is despite ARPES results for the $\mathbf{k}\mathbf{l}$ plane at $\approx20$~K indicating Dirac-like linear bands near $E_{\text{F}}$ \cite{Soh_2019}.

Finally, quantum oscillations have been observed for tetragonal SrMnSb$_2$ and Sr$_{1-x}$K$_x$MnSb$_2$, which have a C-type AFM structure of the Mn moments below $T_{\text{N}}\approx300$~K  \cite{Liu_2017,Liu_2019}, similar to what we find for the Mn sublattice in EuMnSb$_2$. ARPES data for SrMnSb$_2$ are consistent with the size of the Fermi-surface pockets determined from the quantum oscillations \cite{Ramankutty_2018}.  However, DFT and tight-binding calculations find that the Dirac-cone nodes are $\approx200$--$300$~meV above $E_{\text{F}}$ \cite{Ramankutty_2018, Zhang_2019}.   The presence of vacancies at the Sr and Mn crystallographic sites  (i.e.\ Sr$_{1-y}$Mn$_{1-z}$Sb$_2$) have been shown to affect the electronic-band structure \cite{Liu_2017,Saadi_2021}, potentially shifting $E_{\text{F}}$ towards the Dirac nodes \cite{Zhang_2019}.   Thus, these results further question the structural and magnetic properties necessary for tunable topological properties.

Total energy and DFT calculations have shown that  the square geometry of the Sb in SrMnSb$_2$ and the Bi in EuMnBi$_2$ in the $X2$ layers are more amenable to the occurrence of topological fermions than the zigzag chain of the Sb in the $X2$ layers in EuMnSb$_2$ \cite{Yi_2017,Lee_2013,Farhan_2014}. On the other hand, it is also recognized that the details of any magnetic order can also play a role in tuning topological properties \cite{Lee_2013, Farhan_2014}, and our results detailing the various AFM states are important inputs for future calculations.  In particular DFT calculations and complimentary ARPES measurements for the $T<$~\TEut\ phase of EuMnSb$_2$ should be enlightening. 
 
\section{Conclusion}
We have presented a detailed study of the magnetic structures present in the zero-field AFM phases of EuMnSb$_2$.  In addition to the two higher-$T$ phases already observed in some single-crystal studies, we observe a third AFM phase below \TEut$=9(1)$~K.  To summarize, we find that upon cooling below \TMn~$=323(1)$~K, the Mn magnetic sublattice orders into the C-type AFM structure shown in Fig.~\ref{Fig:crys_struct}(e), with $\mu_{\text{Mn}}=3.0(2)~\mu_{\text{B}}$ lying along $\mathbf{a}$ at $T=50$~K.  Below \TEuo$=23(1)$~K, the Eu lattice orders as shown in  Fig.~\ref{Fig:crys_struct}(f) with $\bm{\mu}_{\text{Eu}}$ canted in the $\mathbf{ac}$ plane.  At $12$~K, $\mu_{\text{Eu}}=4.7(2)~\mu_{\text{B}}$ and $\phi_{\text{ac}}=31(1)\degree$.  Finally, below \TEut, $\bm{\mu}_{\text{Eu}}$ develops a large component along $\mathbf{b}$, and $\mu_{\text{Eu}}=5.6(4)~\mu_{\text{B}}$ with components of $2.4(4)$, $4.3(2)$, and $2.6(2)~\mu_{\text{B}}$ lying along $\mathbf{a}$, $\mathbf{b}$, and $\mathbf{c}$, respectively, at $5$~K.  This structure is shown in Fig.~\ref{Fig:crys_struct}(g). The presence of short-range magnetic correlations between $12~\text{K} \alt T \alt 30~\text{K}$ is additionally indicated by the M\"ossbauer, neutron diffraction, and $C_{\text{p}}(T)$ data.  Using these results, we have postulated the signs of NN intraplane and interplane exchange and magnetic anisotropy present within a general Heisenberg-model description of the magnetism.  The signs of these interactions should provide vital input for DFT and other calculations.

We also have discussed how the discovery of magnetic order producing a  net magnetization which in turn gives rise to an effective magnetic field across the Sb$2$ layers would be a highly desirable result.  Whereas we do not find  that this is the case for any of the three AFM states, the discovery of multiple AFM transitions with differing orientations of $\bm{\mu_{\text{Eu}}}$ points to competition between different states and may lead to useful tunability of the magnetism and, in turn, insight into magnetically tuning topological properties in related compounds.  New DFT and total energy calculations incorporating our detailed description for the various AFM states as well as systematic studies subtly varying the amount of disorder in EuMnSb$_2$ should give enlightening results concerning  the topological tunability of $A$Mn$X_2$ compounds.

\begin{acknowledgments}
	 We are grateful for J. Schmidt's assistance with the EDS measurements and appreciate assistance and advice from D.~Vaknin.  This research was supported by the Center for Advancement of Topological Semimetals, an Energy Frontier Research Center funded by the U.S.\ Department of Energy Office of Science, Office of Basic Energy Sciences, through the Ames Laboratory under Contract No.\ DE-AC$02$-$07$CH$11358$. The research was primarily performed at the Ames Laboratory which is operated for the U.S. Department of Energy by Iowa State University under Contract No. DE-AC02-07CH11358.  JMW, YL, SLB, and AK were supported by  Field Work Proposals at the Ames Laboratory. A portion of this research used resources at the Spallation Neutron Source, a U.\,S.\ DOE Office of Science User Facility operated by the Oak Ridge National Laboratory.  Financial support for this work was provided by Fonds Qu\'eb\'ecois de la Recherche sur la Nature et les Technologies, and the Natural Sciences and Engineering Research Council (NSERC) Canada. Much of this work was carried out while DHR was on sabbatical at Iowa State University and their generous support during this visit is gratefully acknowledged.
\end{acknowledgments}

\bibliography{EuMnSb2_diffraction.bib}

\end{document}